%% file: sebob.tex
\documentclass[%
 reprint,
 amsmath,amssymb,
 aps,
 prd,
]{revtex4-2}

\usepackage{graphicx}
\usepackage{dcolumn}
\usepackage{bm}
\usepackage{hyperref}
\usepackage{float}
\usepackage[usenames,dvipsnames]{color}
\usepackage{caption}
\usepackage{subcaption}
\usepackage{comment}
\usepackage[table]{xcolor}
\usepackage{fancyvrb}

\begin{document}


\title{Spinning Effective-to-Backwards One Body (\texttt{SEBOB}): combining Effective One-Body inspirals and Backwards One-Body merger-ringdowns for aligned spin black hole binaries}

\author{Siddharth Mahesh}
\email{sm0193@mix.wvu.edu}
\affiliation{%
    Department of Physics and Astronomy,\\
    West Virginia University
}%
\altaffiliation{Center for Gravitational Waves and Cosmology,\\ West Virginia University}

\author{Sean T. McWilliams}
\email{Sean.McWilliams@mail.wvu.edu}
\affiliation{%
    Department of Physics and Astronomy,\\
    West Virginia University
}%
\altaffiliation{Center for Gravitational Waves and Cosmology,\\ West Virginia University}
\author{Zachariah Etienne}
\affiliation{%
    Department of Physics\\
    University of Idaho
}%

\date{\today}

\begin{abstract}
    High-fidelity gravitational waveform models are essential for realizing the scientific potential of next-generation gravitational-wave observatories. While highly accurate, state-of-the-art models often rely on extensive phenomenological calibrations to numerical relativity (NR) simulations for the late-inspiral and merger phases, which can limit physical insight and extrapolation to regions where NR data is sparse. To address this, we introduce the Spinning Effective-to-Backwards One Body (SEBOB) formalism, a hybrid approach that combines the well-established Effective-One-Body (EOB) framework with the analytically-driven Backwards-One-Body (BOB) model, which describes the merger-ringdown from first principles as a perturbation of the final remnant black hole. We present two variants building on the state-of-the-art \texttt{SEOBNRv5HM} model: \texttt{seobnrv5\_nrnqc\_bob}, which retains standard NR-calibrated non-quasi-circular (NQC) corrections and attaches a BOB-based merger-ringdown; and a more ambitious variant, \texttt{seobnrv5\_bob}, which uses BOB to also inform the NQC corrections, thereby reducing reliance on NR fitting and enabling higher-order ($\mathcal{C}^2$) continuity by construction. Implemented in the open-source \texttt{NRPy} framework for optimized C-code generation, the SEBOB model is transparent, extensible, and computationally efficient. By comparing our waveforms to a large catalog of NR simulations, we demonstrate that SEBOB yields accuracies comparable to the highly-calibrated \texttt{SEOBNRv5HM} model, providing a viable pathway towards more physically motivated and robust waveform models for precision gravitational-wave astronomy.
\end{abstract}

\maketitle

\input{Sections/intro.tex}

\input{Sections/eob.tex}

\input{Sections/bob.tex}

\input{Sections/nrpy.tex}

\input{Sections/results.tex}

\input{Sections/conclusions.tex}

\begin{acknowledgments}
We would like to acknowledge Anuj Kankani, Matthew Cerep, Lorenzo Pompili, Rossella Gamba, and Aasim Jan for useful comments and discussion. Scientific libraries \texttt{numpy}~\cite{vanderWalt:2011bqk} and \texttt{scipy}~\cite{Virtanen:2019joe} were used for part of the accuracy analysis. SM and STM were supported in part by National Science Foundation CAREER grant 1945130 and NASA grants 22-LPS22-0022 and 24-2024EPSCoR-0010. We acknowledge the computational resources provided by the WVU Research Computing Spruce Knob HPC cluster, which is funded in part by NSF grant EPS-1003907, and the Thorny Flat HPC cluster, which is funded in part by NSF grant OAC-1726534. This work was performed in part at the Aspen Center for Physics, which is supported by National Science Foundation grant PHY-2210452.
\end{acknowledgments}

\bibliography{sebob}

\end{document}

%% file: Sections/intro.tex
\section{\label{sec:intro}Introduction}


The last decade of gravitational-wave (GW) astronomy, driven by observations from the LIGO and Virgo detectors, has produced a catalog of $\mathcal{O}(100)$ compact binary coalescences (CBCs)~\cite{KAGRA:2021vkt}. This has inaugurated an era of gravitational-wave astrophysics, yielding unprecedented insights into the populations and properties of stellar remnants, enabling novel tests of General Relativity~\cite{Yunes:2013dva}, new measurements of the cosmic expansion rate~\cite{Palmese:2021mjm}, and a deeper understanding of compact object populations~\cite{LIGOScientific:2020kqk}. Looking ahead, next-generation observatories such as the space-based \textit{Laser Interferometer Space Antenna} (LISA)~\cite{amaro2017laser}, and the planned third-generation ground-based detectors \textit{Einstein Telescope} (ET)~\cite{Punturo:2010zz} and \textit{Cosmic Explorer} (CE)~\cite{Reitze:2019iox}, promise order-of-magnitude improvements in sensitivity. These instruments will regularly produce signals with signal-to-noise ratios (SNRs) in the hundreds or thousands, demanding GW waveform models of correspondingly higher fidelity. Indeed, even at current detector sensitivities, systematic errors in waveform modeling can appreciably bias inferred source parameters~\cite{Kapil:2024zdn}, and reducing such modeling errors will be absolutely critical for future high-SNR science.

The increasing sensitivity and anticipated event rates necessitate highly accurate and computationally efficient gravitational waveform models. These models are the cornerstone of GW astronomy, essential for detecting signals, accurately estimating source parameters, and performing robust astrophysical inference~\cite{Sathyaprakash:2009xt}. Current data analysis pipelines, particularly those based on Bayesian inference, often require millions of waveform evaluations for a single event, demanding models that are both precise and fast.

Several approaches model CBC waveforms. Numerical Relativity (NR) simulations provide the most accurate solutions to Einstein's equations for the strong-field merger phase~\cite{Boyle:2019kee}, serving as crucial benchmarks. Post-Newtonian (PN) theory offers analytical approximations for the early inspiral~\cite{Will:2016sgx} but breaks down near merger. Phenomenological models~\cite{Garcia-Quiros:2020qpx} and surrogate models~\cite{Varma:2018mmi} combine analytical insights with fits to NR data. The Effective-One-Body (EOB) formalism~\cite{Buonanno:1998gg} maps the two-body problem to an effective particle in a deformed metric, incorporating PN information, resummation techniques, and NR calibration. State-of-the-art EOB models like \texttt{SEOBNRv5HM}~\cite{Pompili:2023tna} (which includes higher-order modes and gravitational self-force information) and \texttt{TEOBResumS}~\cite{Nagar:2020pcj} are widely used. While the ultimate goal includes precessing spins (e.g., via \texttt{SEOBNRv5PHM} development), accurate aligned-spin models are a crucial foundation.

This paper focuses on advancing aligned-spin EOB models. While \texttt{SEOBNRv5HM}~\cite{Pompili:2023tna} is highly accurate, its merger-ringdown and non-quasi-circular (NQC) corrections often rely on phenomenological fits to NR data (see Table~\ref{tab:nr_dof}). This empirical tuning can obscure underlying physics and limit extrapolation, especially to regions like extreme mass ratios or nearly extremal spins where NR calibration is sparse.

To address this, we combine EOB with the Backwards-One-Body (BOB) formalism~\cite{McWilliams:2018ztb}. BOB describes the late inspiral, merger, and ringdown by working backwards from the final remnant black hole, effectively treating the late-stage binary as a linear perturbation of the stationary remnant. It leverages the idea that outgoing GWs are driven by perturbations near the remnant's light ring~\cite{Newman:1961qr}, with an underlying orbital frequency smoothly transitioning to the remnant's quasinormal mode (QNM) frequencies~\cite{Berti:2009kk,Vishveshwara:1970zz,Press:1971wr}. Notably, the original BOB model reproduced NR merger-ringdown waveforms to within NR uncertainty \emph{without} phenomenological tuning parameters, suggesting an accurate analytic description of merger physics from first principles.

Our primary contribution is the Spinning Effective-to-Backwards One Body (\texttt{SEBOB}) model. We present two distinct strategies for this integration. The first, termed \texttt{seobnrv5\_nrnqc\_bob}, uses the standard NR-calibrated Non-Quasi-Circular (NQC) corrections from \texttt{SEOBNRv5HM} for the inspiral-plunge phase, then smoothly attaches a BOB model to analytically model the plunge, merger, and ringdown. The second, more ambitious strategy, \texttt{seobnrv5\_bob}, utilizes BOB not only for the merger-ringdown but also to inform the NQC corrections themselves, thereby reducing reliance on direct NR fits and facilitating, by construction, higher-order continuity (e.g., $\mathcal{C}^2$) at the transition point. \texttt{SEBOB} aims to be an analytically driven model for the most nonlinear stage, enhancing robustness where NR calibrations are unreliable. In comprehensive comparisons to SXS NR waveforms (focusing on the dominant $(\ell,m)=(2,2)$ mode), both \texttt{SEBOB} variants achieve median noise-free mismatches of $\mathcal{M}\simeq 2\times 10^{-4}$, competitive with \texttt{SEOBNRv5} and \texttt{TEOBResumS}, even as the BOB-informed variant reduces direct NR calibration (see Sec.~\ref{sec:results}). 

We implement \texttt{SEBOB} within \texttt{NRPy}~\cite{Ruchlin:2017com}, an open-source Python package for symbolic manipulation and optimized C-code generation. \texttt{NRPy} translates EOB Hamiltonians, fluxes, and waveform quantities from a human-readable symbolic form into highly efficient, automatically generated C code, using advanced common subexpression elimination (CSE) on Hamiltonian, flux, and strain expressions. This offers the transparency and flexibility advantages of developing in Python (making it an `experimental playground'), while retaining the performance benefits of C. In wall-time benchmarks (Sec.~\ref{sec:results}), our \texttt{NRPy}-based C implementation yields $\sim3\times$ speedups over \texttt{pySEOBNR}~\cite{Mihaylov:2023bkc} for equal and moderate mass ratios, while remaining competitive with other state-of-the-art generators, including \texttt{lalsuite}'s C-based \texttt{SEOBNRv5\_ROM}s~\cite{lalsuite}(when taking into account the ROM is already in frequency domain). Our analysis also identifies the main sources of residual error near the peak amplitude, pointing to improvements in the treatment of amplitude curvature that we defer to future work.

In summary, \texttt{SEBOB} addresses several needs for next-generation gravitational-wave modeling. It aims for improved \textbf{accuracy} in the late merger-ringdown through analytic insight, which reduces dependence on empirical \textbf{calibration} to NR. Furthermore, it is designed for computational \textbf{efficiency} and scalability suitable for the data rates and parameter-space demands of future observatories. Finally, it promotes enhanced \textbf{reproducibility} and transparency through its open-source implementation. The remainder of this paper is organized as follows: Sec.~\ref{sec:eob} details the \texttt{SEOBNRv5HM} formalism. Sec.~\ref{sec:bob} introduces BOB and our \texttt{SEBOB} integration prescriptions. Sec.~\ref{sec:nrpy} discusses the \texttt{NRPy} implementation. Sec.~\ref{sec:results} presents accuracy and performance comparisons. Sec.~\ref{sec:conclusion} concludes with a discussion of the results and future directions.

Throughout this paper, we employ geometrized units ($c=G=M=1$). Time derivatives are denoted by an overdot (e.g., $\dot{X} \equiv dX/dt$). The symmetric mass ratio of a binary is given by $\eta = m_1 m_2 / (m_1 + m_2)^2$, where $m_1$ and $m_2$ are the masses of the component black holes. The dimensionless spin angular momenta of the individual black holes are represented by $\chi_{i} \equiv \vec{S}_{i}/m_{i}^2$. Waveform angular frequency i.e, derivative of the waveform phase, is denoted by $\omega$ and orbital frequency i.e, derivatives of the azimuthal angle, is denoted by $\Omega$.

%% file: Sections/eob.tex
\section{\label{sec:eob}Spinning EOB Waveforms}

The original EOB formalism \cite{Buonanno:1998gg} canonically mapped the non-spinning, second-order PN two-body problem to the geodesic motion of an effective test particle in a modified Schwarzschild spacetime. Subsequent advances, including higher PN orders, spins, resummation techniques, and calibration against NR data, have yielded state-of-the-art models such as \texttt{SEOBNRv5HM}\cite{Pompili:2023tna} and \texttt{TEOBResumS}\cite{Nagar:2023zxh}. While these models share foundational EOB principles, this work focuses specifically on \texttt{SEOBNRv5HM}. This section details the analytical foundations of \texttt{SEOBNRv5HM}, discusses the incorporation of NR data into the formalism, and outlines the NR parameters essential for constructing inspiral-merger-ringdown (IMR) waveforms.

Within the EOB formalism, the conservative Hamiltonian $\mathcal{H}_\mathrm{real}$ of the PN dynamics is canonically mapped to an effective Hamiltonian $\mathcal{H}_\mathrm{eff}$ via the relation:
\begin{equation}\label{eqn:real_hamiltonian}
    \mathcal{H}_\mathrm{real} = \frac{1}{\eta} \sqrt{1 + 2\eta\left(\mathcal{H}_\mathrm{eff} - 1\right)}.
\end{equation}
The effective Hamiltonian $\mathcal{H}_\mathrm{eff}$ describes the geodesic motion of an effective test particle in a modified Kerr metric $g_{\mu\nu}(x^j)$, with canonical coordinates and momenta denoted by $\{x^j,p_j\}$. It is derived from the relativistic mass-shell constraint, $g^{\mu\nu}p_{\mu}p_{\nu} = -1$, supplemented by a potential $\mathcal{Q}(x^j,p_j)$ that accounts for quartic ($\mathcal{O}(p_j^4)$) and higher-order corrections to geodesic motion. The effective Hamiltonian is given by:
\begin{equation}
    g^{00}\mathcal{H}_\mathrm{eff}^{2} + g^{0i}\mathcal{H}_\mathrm{eff} p_{i} + g^{ij}p_{i}p_{j} + \mathcal{Q}(x^j,p_j) = -1.
\end{equation}
Expressed in spherical polar coordinates $(r,\phi)$ with associated momenta $(p_{r},p_{\phi})$, the effective Hamiltonian also takes the form:
\begin{equation}\label{eqn:effective_hamiltonian}
    \mathcal{H}_\mathrm{eff} = H_\mathrm{odd} + H_\mathrm{even},
\end{equation}
where $H_\mathrm{odd}$ encodes spin-orbit interactions, $H_\mathrm{even}$ describes orbital motion, and $p_\phi$ represents the system's canonical angular momentum. For a detailed expansion of the Hamiltonian, readers are referred to Section II and Appendix A of Ref.~\cite{Pompili:2023tna}.
In the non-spinning limit ($a_i \rightarrow 0$, denoted by the subscript $\mathrm{noS}$), the orbital component $H_\mathrm{even}$ simplifies to:
\begin{equation}
    H_\mathrm{even}|_{a_i \rightarrow 0} = \sqrt{A_\mathrm{noS}\left(1 + A_\mathrm{noS}D_\mathrm{noS}p_{r_*}^2 + \frac{p_{\phi}^{2}}{r^{2}} + Q_\mathrm{noS}\right)}.
\end{equation}
The functions $A_\mathrm{noS}$, $D_\mathrm{noS}$, and $Q_\mathrm{noS}$ incorporate the canonically mapped PN information in this limit. To improve computational efficiency and numerical stability closer to the effective horizon, the coordinates $(r,p_r)$ are transformed to tortoise coordinates $(r_{*},p_{r_{*}})$ via:
\begin{equation}
    \xi = A_\mathrm{noS}\sqrt{D_\mathrm{noS}} ,\quad p_{r_{*}} = \xi p_{r} ,\quad \frac{dr_{*}}{dr} = \frac{1}{\xi}.
\end{equation}

Inspiral gravitational waveform modes $h_{lm}$ are constructed using a factorized representation, $h^\mathrm{F}_{lm}$, derived from PN waveform modes. These modes depend on the binary parameters ($\eta, \chi_1, \chi_2$), the canonical phase space variables $(\vec{r}, \vec{p})$, the EOB Hamiltonian $\mathcal{H}_\mathrm{real}$, the instantaneous orbital frequency $\Omega = d\phi/dt$, and $\Omega_\mathrm{circ}$, which is the corresponding frequency for a circular orbit ($p_{r_{*}} \rightarrow 0$) at radius $r$. Gravitational radiation reaction is incorporated by resumming factorized modes $h^\mathrm{F}_{lm}$ into a dissipative flux term that modify Hamilton's equations. The angular momentum flux $\mathcal{F}_{\phi}$ and the corresponding radial flux $\mathcal{F}_{r}$ are given by:
\begin{align}\label{eqn:factorized_flux}
    \mathcal{F}_{\phi} &= -\frac{\Omega}{8\pi}\sum_{l=2}^{8}\sum_{m=1}^{l} m^2\left|h^\mathrm{F}_{lm}\right|^2,\\
    \mathcal{F}_{r} &= \frac{p_{r_{*}}}{p_\phi}\mathcal{F}_{\phi}.
\end{align}
The canonical coordinates and momenta are evolved numerically using Hamilton's equations of motion, with the fluxes $\mathcal{F}_{\phi}$ and $\mathcal{F}_{r}$ included as external force terms:
\begin{align}
    &\frac{d r}{d t} = \xi\frac{\partial \mathcal{H}_\mathrm{real}}{\partial p_{r_{*}}},\\
    &\frac{d \phi}{d t} = \frac{\partial \mathcal{H}_\mathrm{real}}{\partial p_{\phi}} \equiv \Omega,\\
    &\frac{d p_{r_{*}}}{d t} = -\xi\frac{\partial \mathcal{H}_\mathrm{real}}{\partial r} + \mathcal{F}_{r},\\
    &\frac{d p_\phi}{d t} = \mathcal{F}_{\phi}.
\end{align}
The equations of motion are typically integrated until the orbital frequency $\Omega$ or the time derivative of the radial momentum $dp_{r_{*}}/dt$ reaches an extremum. At each integration timestep, auxiliary quantities ($\mathcal{H}_\mathrm{real}$, $\Omega$, and $\Omega_\mathrm{circ}$) that are necessary for computing the factorized waveform modes $h^\mathrm{F}_{lm}$ are evaluated alongside the canonical variables.

\subsection{Numerical relativity in the EOB formalism}

The EOB formalism, as presented thus far, relies primarily on analytical PN approximations of general relativity. To enhance the accuracy of inspiral waveforms for gravitational-wave signal parameter estimation, corrections from NR are essential. This subsection details the three primary NR inputs used to construct the full IMR waveform: non-quasi-circular (NQC) corrections, the merger-ringdown (MR) waveform model, and inspiral calibration parameters.

First, NQC corrections are applied to improve waveform accuracy in the \emph{plunge regime}, where the quasi-circular orbit approximation used for $h^\mathrm{F}_{lm}$ becomes invalid. After obtaining the EOB trajectory and factorized waveform, the NQC-corrected inspiral-plunge waveform $h^\mathrm{insp-plunge}_{22}$ is computed as:
\begin{multline}\label{eqn:nqc_waveform}
    h^\mathrm{insp-plunge}_{22} = h^\mathrm{F}_{22}\left[1 + \frac{p_{r_{*}}^{2}}{r^2\Omega^2}\left(a_1 + \frac{a_2}{r} + \frac{a_3}{r^{3/2}}\right)\right]\times\\
    e^{i\left(b_1\frac{p_{r_{*}}}{r\Omega} + b_2\frac{p_{r_{*}}^{3}}{r\Omega}\right)}.
\end{multline}
Coefficients $\{a_1,a_2,a_3\}$ and $\{b_1,b_2\}$ are determined by matching $h^\mathrm{insp-plunge}_{22}$'s amplitude and frequency (defined as the phase derivative) to corresponding NR values at a specific time, $t_{22}^{\mathrm{peak}}$. This $t_{22}^{\mathrm{peak}}$ corresponds to the peak amplitude of the NR $(2,2)$ mode strain, related to EOB dynamics via the tunable parameter $\Delta t^\mathrm{ISCO}$:
\begin{equation}
 t_{22}^{\mathrm{peak}} = t^\mathrm{ISCO} - \Delta t^\mathrm{ISCO}.
\end{equation}
Here, $t^\mathrm{ISCO}$ denotes the time the EOB trajectory crosses the innermost stable circular orbit (ISCO) radius of the remnant black hole. The remnant's mass $M_f$ and dimensionless spin $a_f$, essential for ISCO radius calculation, are obtained from NR fitting formulae \cite{Jimenez-Forteza:2016oae,Hofmann:2016yih}.
To determine the amplitude coefficients $\{a_1, a_2, a_3\}$, the factorized EOB waveform and trajectory are interpolated near $t_{22}^{\mathrm{peak}}$. These coefficients are then found by enforcing agreement between the NQC-corrected EOB amplitude $|h^\mathrm{insp-plunge}_{22}|$ (and its first two time derivatives) and the corresponding NR amplitude $|h_{22}^\mathrm{NR}|$ (and its first two time derivatives, with the first derivative being zero at peak amplitude), all evaluated at $t_{22}^{\mathrm{peak}}$:
\begin{gather}\label{eqn:nqc_corrections_amplitude}
    \begin{pmatrix}
    \mathcal{Q}_{1} & \mathcal{Q}_{2} & \mathcal{Q}_{3} \\
    \dot{\mathcal{Q}}_{1} & \dot{\mathcal{Q}}_{2} & \dot{\mathcal{Q}}_{3} \\
    \ddot{\mathcal{Q}}_{1} & \ddot{\mathcal{Q}}_{2} & \ddot{\mathcal{Q}}_{3}
    \end{pmatrix}
    \begin{pmatrix}
        a_1 \\ a_2 \\ a_3
    \end{pmatrix}
    =
     \begin{pmatrix}
        |h_{22}^\mathrm{NR}| - |h_{22}^\mathrm{F}| \\
        \tfrac{d}{dt}|h_{22}^\mathrm{NR}| - \tfrac{d}{dt}|h_{22}^\mathrm{F}| \\
        \tfrac{d^2}{dt^2}|h_{22}^\mathrm{NR}| - \tfrac{d^2}{dt^2}|h_{22}^\mathrm{F}|
    \end{pmatrix}_{t_{22}^\mathrm{peak}},
    \\
    \mathcal{Q} = \left[|h^\mathrm{F}_{22}|\left\{\frac{p_{r_{*}}^{2}}{r^2\Omega^2},\frac{p_{r_{*}}^{2}}{r^3\Omega^2},\frac{p_{r_{*}}^{2}}{r^{7/2}\Omega^2}\right\}\right]_{t = t^\mathrm{peak}_{22}}.
\end{gather}
Similarly, phase correction coefficients $\{b_1, b_2\}$ are determined by matching the NQC-corrected frequency $\omega_{22} = d\phi_{22}/dt$ and its time derivative $\dot{\omega}_{22}$ of the NQC-corrected EOB waveform to corresponding NR values at $t_{22}^{\mathrm{peak}}$:
\begin{gather}\label{eqn:nqc_corrections_frequency}
    \begin{pmatrix}
    \mathcal{P}_{1} & \mathcal{P}_{2}\\
    \dot{\mathcal{P}}_{1} & \dot{\mathcal{P}}_{2}\\
    \end{pmatrix}
    \begin{pmatrix}
        b_1 \\ b_2
    \end{pmatrix}
    =
     \begin{pmatrix}
        \omega_{22}^\mathrm{NR} - \omega_{22}^\mathrm{F} \\
        \dot{\omega}_{22}^\mathrm{NR} - \dot{\omega}_{22}^\mathrm{F} \\
    \end{pmatrix}_{t_{22}^\mathrm{peak}},
    \\
    \mathcal{P} = \left[\left\{\frac{p_{r_{*}}}{r\Omega},\frac{p_{r_{*}}^{3}}{r\Omega}\right\}\right]_{t = t^\mathrm{peak}_{22}}.
\end{gather}

Given the trajectory and the factorized waveform, the quantities $\mathcal{Q}$ and $\mathcal{P}$ can be interpolated, spline-differentiated, and evaluated at $t_{22}^\mathrm{peak}$. Equations \ref{eqn:nqc_corrections_amplitude} and \ref{eqn:nqc_corrections_frequency} are then solved numerically to determine the NQC coefficients $\{a_1, a_2, a_3\}$ and $\{b_1, b_2\}$ respectively.
Second, the EOB formalism incorporates a separate MR model to specify the full gravitational strain of a black hole binary merger. In \texttt{SEOBNRv5}, MR waveform modes are modeled as the emission of \emph{quasi-normal modes} (QNMs)\cite{Berti:2009kk} from the remnant black hole, with amplitudes and phases modulated by phenomenological functions fitted to NR data:
\begin{gather}\label{eqn:seobnrv5_mr_waveform}
    h^\mathrm{merger-ringdown}_{22}(t) = \tilde{A}_{22}(t) e^{i\tilde{\phi}_{22}(t)}e^{-i\sigma^\mathrm{QNM}_{22}(t - t_{22}^\mathrm{peak})},\\
    \tilde{A}_{22}(t) = \eta c_{1,c}\mathrm{tanh}\left[c_{1,f}\left(t - t_{22}^\mathrm{peak}\right) + c_{2,f}\right] + c_{2,c},\\
    \tilde{\phi}_{22} = \phi_{0} - d_{1,c}\mathrm{log}\left[\frac{1 + d_{2,f}e^{-d_{1,f}\left(t - t^\mathrm{peak}_{22}\right)}}{1 + d_{2,f}}\right],\\
    \sigma^\mathrm{QNM}_{22} = -\frac{i}{\tau^\mathrm{QNM}_{22}} + \omega_{22}^\mathrm{QNM}.
\end{gather}
Here, $\omega_{22}^\mathrm{QNM}$ and $\tau_{22}^\mathrm{QNM}$ are the frequency and damping time of the dominant $(2,2)$ QNM. These are calculated as functions of the remnant mass $M_f$ and spin $a_f$ (obtained from NR fits) using tools such as the \texttt{qnm} package \cite{Stein:2019mop}. Coefficients $\{c_{1,f}, c_{2,f}, d_{1,f}, d_{2,f}\}$ are determined by fitting to a catalog of NR waveforms. The remaining coefficients $\{c_{1,c}, c_{2,c}, d_{1,c}, \phi_{0}\}$ are fixed by requiring $C^1$ continuity (continuous value and first derivative) when attaching the MR waveform to the NQC-corrected inspiral-plunge waveform at $t_{22}^{\mathrm{peak}}$.

Third, after NQC corrections and the MR model structure are incorporated, the overall accuracy of the resulting IMR waveform is further enhanced by tuning three parameters $\left\{a_6,d_\mathrm{SO},\Delta t^\mathrm{ISCO}\right\}$ within the EOB dynamics against a large set of NR waveforms. The parameter $a_6$ is a pseudo-PN coefficient entering the non-spinning inspiral potential $A_\mathrm{noS}$ at 5PN order:
\begin{multline}
    A_\mathrm{noS} = 1 - 2u + 2\eta u^{3} \\
    + \eta\left(\frac{94}{3} - \frac{41\pi^2}{32}\right) u^4 \\
    + \Bigg[ \eta\left(\frac{2275\pi^2}{512} - \frac{4237}{60} + \frac{128\gamma_\mathrm{E}}{5} + \frac{256\ln 2}{5}\right)\\
    + \eta^2\left(\frac{41\pi^2}{32} - \frac{221}{6}\right)\\
    + \frac{64}{5}\eta\ln u\Bigg]u^5\\
    + \left[\eta a_6 + \left(-\frac{144\eta^2}{5} - \frac{7004\eta}{105}\right)\ln u\right]u^6,
\end{multline}
where $u \equiv 1/r$ is the inverse of the radial separation. The parameter $d_\mathrm{SO}$ is a pseudo-PN parameter that enters the spin-orbit Hamiltonian at the 4.5PN order:
\begin{equation}
    H_\mathrm{odd} = \frac{p_\phi\left(g_{a_{+}}a_{+} + g_{a_{-}}a_{-} \right) + \eta d_\mathrm{SO}u^{3} p_\phi a_{+} + G^\mathrm{align}_{a^3}}{r^3 + a_{+}^{2}\left(r + 2\right)},
\end{equation}
where the spin combinations $a_{\pm} = \chi_1m_1 \pm \chi_2m_2$, $g_{a_{\pm}}$ are the corresponding gyrogravitomagnetic factors, and $G^\mathrm{align}_{a^3}$ contains cubic-in-spin couplings (See Equations(26)-(30) of ~\cite{Khalil:2023kep}). The third $\Delta t^\mathrm{ISCO}$ parameter, defined earlier, encodes the time between ISCO crossing and the peak of the NR waveform. These three parameters are calibrated via Nested Bayesian inference, minimizing mismatch against a set of 446 SXS \cite{Boyle:2019kee} (and one Einstein Toolkit) waveforms. The inference proceeds in stages: first, non-spinning waveforms constrain $a_6$ and the spin-independent components of $\Delta t^\mathrm{ISCO}$; subsequently, $d_\mathrm{SO}$ and the spin-dependent components of $\Delta t^\mathrm{ISCO}$ are constrained. Crucially, this calibration is performed on NQC-corrected, merger-attached waveforms. Consequently, NR-derived parameters for NQC corrections ($|h_{22}^\mathrm{NR}|^\mathrm{peak}, \ddot{|h|}_{22}^\mathrm{peak}, \omega_{22}^\mathrm{peak}, \dot{\omega}_{22}^\mathrm{peak}$) and MR model fits ($c_{i,f}, d_{i,f}, M_f, a_f$) are considered ``pre-calibration'' inputs, fixed prior to the final tuning of $\{a_6, d_\mathrm{SO}, \Delta t^\mathrm{ISCO}\}$. The complete set of NR-derived parameters for the \texttt{SEOBNRv5} model is summarized in Table~\ref{tab:nr_dof}.

The \texttt{SEOBNRv5} waveform model, which combines novel PN approximation approaches with advances in accurate NR simulations, provides highly accurate templates for analyzing gravitational wave signals from black hole binary mergers observed by LIGO, Virgo, and KAGRA. The model \cite{Pompili:2023tna} is implemented in \texttt{Cython} within the open-source package \texttt{pySEOBNR}\cite{Mihaylov:2023bkc}. For faster waveform evaluation, a reduced-order surrogate model, \texttt{SEOBNRv5\_ROM}, is also implemented in C as part of \texttt{lalsuite}\cite{lalsuite}.

\begin{table*}
    \caption{\label{tab:nr_dof}Summary of NR inputs for the (2,2) mode of the \texttt{SEOBNRv5} waveform model. The first three parameters are calibration parameters, tuned via Bayesian inference against NR waveforms. The remaining parameters are derived from NR fits and are specified, prior to calibration, for NQC corrections or the merger-ringdown model. Nominal values for these parameters are found in \cite{Pompili:2023tna}, except for $M_f$ and $a_f$, which are given in the listed citations in the Method column. The abbreviation NBI stands for Nested Bayesian Inference.}
    \begin{ruledtabular}
    \begin{tabular}{ccccc}
    \textbf{Parameter}&\textbf{Description}&\textbf{Pre Calibration}&\textbf{Method}&\textbf{In \texttt{seobnrv5\_bob}}\\
    \hline
    $a_6$&Effective 5PN non-spinning inspiral coefficient&-&NBI&\checkmark\\
  $d_\mathrm{SO}$&Effective 4.5PN inspiral spin-orbit coefficient&-&NBI&\checkmark\\
  $\Delta{t}^\mathrm{ISCO}$&Time shift between EOB ISCO passage and (2,2) peak strain&-&NBI&\checkmark\\
  $|h_{22}^\mathrm{NR}|^\mathrm{peak}$&(2,2) mode peak strain amplitude&\checkmark&NR fit&\checkmark\\
  $\ddot{|h|}_{22}^\mathrm{peak}$&(2,2) mode strain amplitude 2nd deriv. at peak strain time&\checkmark&NR fit& -\\
  $\omega_{22}^\mathrm{peak}$&(2,2) mode frequency at peak strain time&\checkmark&NR fit&\checkmark\\
  $\dot{\omega}_{22}^\mathrm{peak}$&(2,2) mode frequency deriv. at peak strain time&\checkmark&NR fit& -\\
  $M_\mathrm{f}$&Remnant Mass&\checkmark&NR fit \cite{Jimenez-Forteza:2016oae}&\checkmark\\
  $a_\mathrm{f}$&Remnant Spin&\checkmark&NR fit \cite{Hofmann:2016yih}&\checkmark\\
  $c_{1,f}$&Merger-ringdown amplitude coefficient&\checkmark&NR fit& -\\
  $c_{2,f}$&Merger-ringdown amplitude coefficient&\checkmark&NR fit& -\\
  $d_{1,f}$&Merger-ringdown phase coefficient&\checkmark&NR fit& -\\
  $d_{2,f}$&Merger-ringdown phase coefficient&\checkmark&NR fit& -\\
  \end{tabular}
\end{ruledtabular}
\end{table*}

%% file: Sections/bob.tex
\section{\label{sec:bob}Backwards One Body Waveforms}
The \emph{Backwards One Body} (BOB) formalism~\cite{McWilliams:2018ztb} models the late inspiral and merger-ringdown stages of black hole binary coalescence. It describes these phases as dynamical perturbations propagating outwards through null congruences near the remnant's light ring, superposed on the stationary spacetime of the merged remnant black hole. This section first reviews BOB's mathematical formulation, then presents and justifies two distinct prescriptions for matching EOB inspiral waveforms. Throughout this section, $\omega(t)$ and related quantities like $\omega_0$ and $\omega_\mathrm{QNM}$ refer to the gravitational waveform frequency for a given $(l,m)$ mode. 

The BOB formalism models gravitational wave perturbation amplitudes using the \emph{Newman-Penrose scalar}~\cite{Newman:1961qr} $\psi_4$, which represents the second derivative of the gravitational wave strain, $\ddot{h}$. Based on the divergence of null congruences near the light ring, the amplitude of each $(l,m)$ mode is given by (cf. Eq.~5 in Ref.~\cite{McWilliams:2018ztb}):
\begin{equation}\label{eqn:bob_amplitude_derivation}
    A_{lm}(t) \equiv |\psi_{4,lm}| = A_{p,lm}\, \mathrm{sech}\left(\frac{t - t_{p,lm}}{\tau_{lm}}\right),
\end{equation}
where $(t_{p,lm},A_{p,lm})$ correspond to the time and value of the peak $\psi_{4,lm}$ amplitude for that mode, and $\tau_{lm}$ is the damping time of the $(l,m)$ \emph{quasinormal mode} (QNM)~\cite{Berti:2009kk} of the remnant black hole.

The waveform phase derivation assumes a single underlying orbital frequency $\Omega(t)$, which governs all multipole frequencies via $\omega_{lm} = m\Omega$. This frequency evolution is inferred from the identity~\cite{Baker:2008mj,Kelly:2011bp} relating the news function, $\mathcal{N}_{lm} \equiv \dot{h}_{lm}$, to the frequency evolution:
\begin{equation}\label{eqn:bob_frequency_derivation}
    |\mathcal{N}_{lm}|^2 \propto m^2\frac{d \Omega^2}{dt}.
\end{equation}
Combined with the adiabatic relation $|\psi_{4,lm}|^2 \approx m^2\Omega^2|\mathcal{N}_{lm}|^2$ (more precisely, $|\psi_{4,lm}|^2 \propto m^4 \Omega^3 \frac{d\Omega}{dt}$ which leads to the form below) that assumes a negligible contribution from the news amplitude derivative to $\psi_4$, we obtain an expression for $\omega_{lm} = m\Omega$. Dropping mode labels, the frequency $\omega(t)$ is given by:
\begin{equation}\label{eqn:bob_frequency}
    \omega(t) = \left\{ \omega_{0}^{4} + k\left(\mathrm{tanh}\left(\tfrac{t - t_p}{\tau}\right) - \mathrm{tanh}\left(\tfrac{t_0 - t_p}{\tau}\right)\right) \right\}^{1/4},
\end{equation}
with $k$ defined as:
\begin{equation}
    k = \frac{\omega_{\mathrm{QNM}}^{4} - \omega_{0}^{4}}{1 - \mathrm{tanh}\left(\tfrac{t_0 - t_p}{\tau}\right)},
\end{equation}
where $t_0$ and $\omega_0$ are the time and waveform frequency at a reference point, and $\omega_{\mathrm{QNM}}$ is the fundamental QNM frequency of the remnant black hole. Integrating $\omega(t)$ then yields the waveform phase $\phi(t)$:
\begin{multline}\label{eqn:bob_phase}
    \phi(t) = \phi_{0} + \omega_{+\infty}\tau\left(\mathrm{arctan}_{+} + \mathrm{arctanh}_{+}\right)\\
      - \omega_{-\infty}\tau\left(\mathrm{arctan}_{-} + \mathrm{arctanh}_{-}\right),
\end{multline}
where $\phi_0$ is the phase at the reference time $t_0$, $\omega_{\pm\infty}$ are the frequencies evaluated in the limit $t \to \pm\infty$, and the terms $\mathrm{arctan}_{\pm}$/$\mathrm{arctanh}_{\pm}$ are defined as:
\begin{align}
    \mathrm{arctan}_{\pm} &= \left( \mathrm{tan}^{-1}\left(\frac{\omega}{\omega_{\pm\infty}}\right) - \mathrm{tan}^{-1}\left(\frac{\omega_0}{\omega_{\pm\infty}}\right) \right),\\
    \mathrm{arctanh}_{\pm} &= \left( \mathrm{tanh}^{-1}\left(\frac{\omega}{\omega_{\pm\infty}}\right) - \mathrm{tanh}^{-1}\left(\frac{\omega_0}{\omega_{\pm\infty}}\right) \right),
\end{align}
where $\mathrm{tan}^{-1}$ and $\mathrm{tanh}^{-1}$ are the inverse tangent and inverse hyperbolic tangent respectively.

To relate $\psi_4$ to the gravitational wave strain $h$, we assume the phase evolves linearly at leading order near merger, yielding $\psi_4 = \ddot{h} \approx -\omega(t)^2 h$. Under this approximation, the complex strain is given by:
\begin{equation}\label{eqn:bob_complex_strain}
    h(t) = \frac{|h_0|\omega_0^2}{\omega(t)^2}\frac{\mathrm{sech}\left(\tfrac{t - t_p}{\tau}\right)}{\mathrm{sech}\left(\tfrac{t_0 - t_p}{\tau}\right)} \exp\left(i\phi(t)\right),
\end{equation}
where $|h_0|$ is the strain amplitude at the reference time $t_0$.

A condition relating $t_p$ to $t_0$ is required. For matching with \texttt{SEOBNRv5}, $t_0$ is chosen as the time of peak strain amplitude, $t^{\mathrm{peak}}_{22}$, the natural end point of the EOB inspiral. Requiring $d|h|/dt = 0$ at $t = t_0$ leads to the following expression:
\begin{equation}\label{eqn:tp}
    t_p = t_0 - 2\tau\ln\left(\frac{\omega_0}{\omega_\mathrm{QNM}}\right),
\end{equation}
where the identity
\begin{equation}
    \tanh^{-1}(x) = \frac{1}{2}\ln\left(\frac{1+x}{1-x}\right)
\end{equation}
is used to invert the hyperbolic tangent function.
Equation \eqref{eqn:tp} differs from the $t_p$ condition derived in Eq.~9 of Ref.~\cite{McWilliams:2018ztb}, which imposes $\mathcal{C}^2$ phase continuity and depends on $\dot{\omega}_0$ for an arbitrary matching time $t_0$. However, within the EOB framework, \emph{non-quasi-circular} (NQC) corrections ensure a peak in the strain at the matching time. This choice for $t_0$ therefore ensures the matching point corresponds to the peak strain amplitude in both the EOB and BOB models. This approach, implemented in the \texttt{seobnrv5\_nrnqc\_bob} model, utilizes BOB to model the merger-ringdown, attached to the NQC-corrected EOB inspiral.

Alternatively, BOB can directly provide NQC corrections. The conventional \texttt{SEOBNRv5} NQC corrections [cf. Eq.~\ref{eqn:nqc_waveform}] rely on NR fits to the peak amplitude, frequency, and their derivatives at the peak strain time. With the BOB approach, these derivatives are instead computed analytically from the BOB model (Eqs.~\ref{eqn:bob_complex_strain} and~\ref{eqn:bob_frequency}) evaluated at the peak strain time. The functional form of the NQC-corrected inspiral-plunge waveform from Eq.~\eqref{eqn:nqc_waveform} is maintained,
but the coefficients $\{a_1,a_2,a_3,b_1,b_2\}$ are instead determined by requiring agreement with the BOB-predicted values for the strain amplitude $|h_\mathrm{BOB}|$, its first time derivative (which is zero at $t_0$ by construction using Eq.~\eqref{eqn:tp}), its second time derivative $d^2|h_\mathrm{BOB}|/dt^2$, as well as the frequency $\omega_\mathrm{BOB}$ and its first time derivative $d\omega_\mathrm{BOB}/dt|$, all evaluated at $t_0$. This model is implemented in NRPy as \texttt{seobnrv5\_bob}.

The approach of using BOB to inform NQCs is motivated by the observation that the peak strain occurs close to the perturber crossing the remnant \emph{innermost stable circular orbit} (ISCO), a regime particularly apt for BOB's description. It reduces reliance on direct NR fitting for NQC parameters by leveraging BOB's analytical structure. Primary NR inputs are thus limited to the remnant properties (mass and spin, which determine $\omega_\mathrm{QNM}$ and $\tau$) and the strain amplitude $|h_0|$ and frequency $\omega_0$ at the matching time $t_0$. Furthermore, while the \texttt{SEOBNRv5} merger-ringdown attachment guarantees only $\mathcal{C}^1$ continuity (continuous complex $h$ and $\dot{h}$) at the matching point $t_0$, employing BOB for NQC computation inherently links inspiral corrections to the merger-ringdown model, thus enabling $\mathcal{C}^2$ continuity (continuous $h, \dot{h}, \ddot{h}$) by construction.

Conversely, NQC corrections are applied to the entire inspiral waveform via their dependence on EOB dynamical variables ($p_{r_*}, r, \Omega$). For generic non-circular orbits, the physically correct dependence may involve \emph{post-Newtonian} (PN) waveform predictions. In such cases, BOB may not provide a complete description of the inspiral's NQC corrections. Therefore, employing BOB solely as a merger-ringdown attachment (\texttt{seobnrv5\_nrnqc\_bob}) retains the NR-calibrated NQCs for the inspiral and could be advantageous.

Equation \eqref{eqn:tp} implies that the merger-ringdown waveform is determined solely by the strain amplitude $|h_0|$ and frequency $\omega_0$ at the matching time $t_0$, along with the remnant mass and spin used to calculate $\omega_\mathrm{QNM}$ and $\tau$. Thus, BOB waveforms eliminate all NR-informed degrees of freedom in modeling the merger-ringdown evolution, provided that the remnant mass and spin themselves are predicted from inspiral parameters without direct NR fitting (e.g., using methods like those in Ref.~\cite{Buonanno:2005xu}). This can also aid in calibrating the EOB Hamiltonian to accurately predict remnant mass and spin directly from the inspiral dynamics (see \cite{Buonanno:2005xu}). Although the BOB formalism can model any gravitational wave mode $(l,m)$, our analysis in this paper is restricted to the dominant $(2,2)$ mode. The combination of BOB's analytical model for the merger-ringdown with the EOB inspiral's analytical foundations into an \emph{Effective-to-Backwards-One-Body} (EBOB) framework aims to describe the complete black hole binary coalescence waveform analytically and from physical motivations. In Section \ref{sec:results}, we will assess the fidelity of BOB predictions against numerical relativity data and examine the impact of using BOB-derived NQCs on overall waveform accuracy.

%% file: Sections/nrpy.tex
\section{\label{sec:nrpy}Numerical Implementation}

The \texttt{SEBOB} waveform model, comprising the \texttt{seobnrv5\_nrnqc\_bob} and \texttt{seobnrv5\_bob} variants, is implemented in the open-source \emph{\texttt{NRPy}} framework~\cite{Ruchlin:2017com,Etienne:2024ncu}. \texttt{NRPy}, built upon the \texttt{SymPy} library~\cite{Meurer:2017yhf} for symbolic manipulation, translates all required EOB and BOB mathematical expressions into highly optimized, modular C code. This symbolic-to-numerical workflow, which extensively applies Common Subexpression Elimination (CSE) to reduce runtime costs, facilitates transparency, extensibility, and high performance.

The generated C code is self-contained and can be compiled as an executable or linked as a library. It relies on the GNU Scientific Library (GSL)~\cite{Galassi:2019czg} for core numerical tasks, including root-finding, ordinary differential equation (ODE) integration, interpolation, and linear algebra. This structure enables seamless integration into gravitational-wave data analysis packages such as \texttt{LALSuite}~\cite{lalsuite} by avoiding Python-to-C overhead, and supports deployment within infrastructures such as BlackHoles@Home~\cite{Etienne:2024ncu} for large-scale model calibration and waveform generation.

\subsection{EOB Dynamics: Initial Conditions and Evolution}

\subsubsection{Initial Conditions}
EOB evolution begins by determining initial conditions for the canonical phase-space variables $\{r, \phi, p_{r_*}, p_{\phi}\}$. First, for a user-specified initial orbital frequency $\Omega_0$, the initial orbital separation $r_0$ and azimuthal momentum $p_{\phi,0}$ for quasi-circular orbits are obtained by solving the two-dimensional non-linear algebraic system:
\begin{align}
    \frac{dp_{r}}{dt} \equiv \frac{\partial \mathcal{H}_\mathrm{real}(r, p_{r_*}=0, p_{\phi})}{\partial r} &= 0, \\
    \frac{d\phi}{dt} \equiv \frac{\partial \mathcal{H}_\mathrm{real}(r, p_{r_*}=0, p_{\phi})}{\partial p_{\phi}} &= \Omega_0.
\end{align}
This system is solved using GSL's hybrid Powell multi-dimensional root-finding algorithm. The first condition is equivalent to no radial motion since $p_{r_{*}} = 0$ guarantees $dr/dt = 0$ since the Hamiltonian is quadratic in $p_{r_{*}}$.

Second, the initial radial momentum $p_{r_{*,0}}$, typically small but non-zero for dissipative inspirals, is determined by solving a one-dimensional root-finding problem using GSL's Brent-Dekker solver. The radial momentum is derived by equating the initial conservative radial velocity $\dot{r}_\mathrm{dyn} = \xi \partial \mathcal{H}_\mathrm{real}/\partial p_{r_*}$ to the radial velocity due to radiation reaction $\dot{r}_\mathrm{rad} = \dot{r}_\mathrm{dyn}$ in the adiabatic limit, with $\dot{r}_\mathrm{rad}$ given by~\cite{Buonanno:2005xu}
\begin{align}
    \dot{r}_\mathrm{dyn} &\equiv \xi \frac{\partial \mathcal{H}_\mathrm{real}(r=r_0,p_{r_*},p_{\phi}=p_{\phi,0})}{\partial p_{r_*}},\\
    \dot{r}_\mathrm{rad} &\equiv \frac{\mathcal{F}_\phi}{{dE}/{dr}},\\
    \frac{dE}{dr} &\equiv -\left[\frac{ \frac{\partial \mathcal{H}}{\partial{p_{\phi}}} \frac{\partial^2\mathcal{H}}{\partial{r^2}} }{ \frac{\partial^2\mathcal{H}}{\partial{p_{\phi}}\partial{r}} }\right]_{\left(r=r_0,p_{r_*},p_{\phi} = p_{\phi,0}\right)} .
\end{align}
The initial phase $\phi_0$ is set to zero.

\subsubsection{Orbital Evolution}
With initial conditions established, Hamilton's equations of motion are integrated forward in time using a GSL adaptive 8th-order Runge-Kutta Prince-Dormand method (or similar) with error control. Tortoise coordinates $(r_*, p_{r_*})$ are used to enhance numerical stability, particularly near the EOB horizon analogue. Integration terminates after a radial separation of $6 M$ is reached \emph{and} one of the following conditions are met:

\begin{itemize}
    \item The orbital frequency $\Omega$ reaches a maximum and begins to decrease, indicating an unphysical peak in frequency evolution.
    \item $dp_{r_*}/dt$ becomes positive or peaks, indicating an unphysical outspiral.
    \item $dr/dt$ becomes positive, indicating an unphysical outspiral.
    \item The orbital separation $r$ drops below a predefined minimum (e.g., related to $r_\mathrm{ISCO}$). This is only realized in cases where the parameter $\Delta t^\mathrm{ISCO}$ is positive, indicating the NR waveform strain peaks before the EOB trajectory reaches the ISCO of the remnant black hole.
    \item The circular orbit frequency $\Omega_\mathrm{circ}$ exceeds unity at radius $r < 3M$. This is indicative of unphysically high rotation speeds and is used to terminate the integration.
\end{itemize}
Throughout the integration, quantities required for radiation reaction fluxes ($\mathcal{F}_r, \mathcal{F}_{\phi}$) and waveform modes ($h^\mathrm{F}_{lm}$), such as $\mathcal{H}_\mathrm{real}$, $\Omega$, and $\Omega_\mathrm{circ}$, are computed from the phase-space state and binary parameters. These quantities are stored at variable resolution, with coarser sampling during the early, slower inspiral and finer sampling of the last $250 M$ in time for performing NQC corrections.

\subsubsection{Factorized Waveform Generation}
Factorized inspiral waveform modes $h^\mathrm{F}_{lm}$ are computed at each step of the finely sampled EOB trajectory. These modes are constructed symbolically within \texttt{NRPy} and generated into optimized C code. This symbolic generation significantly facilitates faster computation of waveform modes, particularly by enabling pre-computation of coefficients that depend only on constant binary parameters. This benefit is particularly important for precessing binaries, where these coefficients become time-dependent due to spin vector evolution and thus cannot be precomputed during integration.

\subsection{Non-Quasi-Circular Corrections and Merger Attachment}
To accurately model the late inspiral and plunge, Non-Quasi-Circular (NQC) corrections are applied to the dominant $(l=2, m=2)$ factorized EOB mode $h^\mathrm{F}_{22}$, as described by Eq.~\eqref{eqn:nqc_waveform}. This involves determining the NQC amplitude coefficients $\{a_1, a_2, a_3\}$ and phase coefficients $\{b_1, b_2\}$ by solving $3 \times 3$ and $2 \times 2$ linear systems (Eqs.~\eqref{eqn:nqc_corrections_amplitude} and \eqref{eqn:nqc_corrections_frequency}), respectively, using GSL routines for LU decomposition and forward/backward substitution.

These systems require quantities computed at the predicted matching time, $t_0$, defined as:
\begin{align}
    t_0 &= t^\mathrm{ISCO} - \Delta t^\mathrm{ISCO},
\end{align}
where $t^\mathrm{ISCO}$ is the time the EOB trajectory crosses the innermost stable circular orbit (ISCO) radius of the remnant black hole, and $\Delta t^\mathrm{ISCO}$ is a calibration parameter that sets the time at which the NR waveform strain peaks relative to the ISCO time. Both $t^\mathrm{ISCO}$ and $t_0$ are determined from the finely sampled portion of the trajectory by identifying the nearest index in the radial or time arrays, followed by interpolation between the 10 nearest sample points at higher resolution.

In a minority of cases, if ODE integration terminates before the perturber crosses the remnant ISCO or before $t_0$, then $t_0$ is set to the last sample point of the trajectory. Consistency in NQC coefficient calculation is crucial, as GSL cubic spline interpolation (used for time derivatives at the last sample point) applies natural boundary conditions, while \texttt{pySEOBNR} (also used) employs not-a-knot boundary conditions, leading to significant differences. To ensure consistency, we modified \texttt{pySEOBNR} to use natural boundary conditions for all its cubic spline interpolations.

\subsubsection{\texttt{seobnrv5\_nrnqc\_bob}: NR-informed NQCs and BOB Merger-Ringdown}
This variant mirrors the \texttt{SEOBNRv5HM} approach for NQC corrections. The target values for the right-hand side of Eqs.~\eqref{eqn:nqc_corrections_amplitude} and \eqref{eqn:nqc_corrections_frequency}---specifically, the amplitude $|h_{22}^\mathrm{NR}|$, its first and second time derivatives, the frequency $\omega^\mathrm{NR}_{22}$, and its first time derivative---are derived from fits to NR simulations, evaluated at the peak of the (2,2) mode of the strain. After constructing the NQC-corrected inspiral-plunge waveform $h^\mathrm{insp-plunge}_{22}$, the BOB model (Sec.~\ref{sec:bob}) is attached at $t_0$ to describe the merger and ringdown. This attachment procedure ensures $\mathcal{C}^1$ continuity of the complex strain $h$ and its first time derivative $\dot{h}$ at $t_0$.

\subsubsection{\texttt{seobnrv5\_bob}: BOB-informed NQCs and Merger-Ringdown}
This model integrates BOB more deeply into the inspiral by leveraging it to inform the NQC corrections. The functional form of the NQC corrections (Eq.~\eqref{eqn:nqc_waveform}) is maintained, but the target values for the linear systems are sourced analytically from the BOB model itself (Eqs.~\eqref{eqn:bob_complex_strain}, \eqref{eqn:bob_frequency}, and \eqref{eqn:tp}), evaluated at $t_0 = t_{22}^\mathrm{peak}$. Specifically, the required amplitude and frequency derivatives at the matching time, ($d^2|h_\mathrm{BOB}|/dt^2|_{t_0}$ and $d\omega_\mathrm{BOB}/dt|_{t_0}$) are computed analytically from the BOB equations, with \texttt{NRPy} performing symbolic differentiation and generating C code for their evaluation. Remnant parameters $M_f$ and $a_f$ (essential for computing $\omega_\mathrm{QNM}$ and $\tau$ within the BOB model) are obtained from NR fits~\cite{Jimenez-Forteza:2016oae,Hofmann:2016yih}. The QNM frequency $\omega_\mathrm{QNM}$ and damping time $\tau$ are computed by interpolating over a set of 107 values of the final spin and the corresponding QNM values computed in Python using the \texttt{qnm} package~\cite{Stein:2019mop}.

Once NQC coefficients are found using either BOB-derived or NR-informed targets, $h^\mathrm{insp-plunge}_{22}$ is constructed by multiplying the factorized waveform $h^\mathrm{F}_{22}$ by the NQC correction given in Eq.~\eqref{eqn:nqc_waveform}. Since the factorized waveform is computed at the sample points of the ODE trajectory, no additional interpolation is required. For $t \ge t_0$, the waveform is subsequently described by the same BOB model used for NQC targets. This construction, leveraging BOB-derived NQC targets and the BOB model for merger-ringdown, inherently ensures $\mathcal{C}^2$ continuity of $h, \dot{h}, \text{and } \ddot{h}$ at $t_0$.

In order to perform consistency checks against the \texttt{SEOBNRv5HM} model, we also implement the native \texttt{SEOBNRv5HM} merger ringdown model given in Eq.~\eqref{eqn:seobnrv5_mr_waveform}. This model is implemented in \texttt{NRPy} as \texttt{seobnrv5\_nrpy}. 

The \texttt{SEBOB} waveform code used in this paper can be generated with the following commands (flag \texttt{-h} provides instructions to generate specific variants):

\begin{widetext}
\begin{Verbatim}[numbers=left]
pip install git+https://github.com/nrpy/nrpy.git@e16a119
python3 -m nrpy.examples.seobnrv5_aligned_spin_inspiral -h
\end{Verbatim}
\end{widetext}
The C code can be built and run with the following commands (flag \texttt{-h} provides instructions to pass input parameters):
\begin{Verbatim}[numbers=left]
cd project/seobnrv5_aligned_spin_inspiral
make        
./seobnrv5_aligned_spin_inspiral -h
\end{Verbatim} 
To access up-to-date versions of developmental features discussed in this paper, we recommend directly cloning the repository -- \texttt{git clone git@github.com:nrpy/nrpy.git}.

%% file: Sections/results.tex
\section{\label{sec:results}Results}

We present a systematic evaluation of the \texttt{SEBOB} waveform model. After verifying numerical consistency between our \texttt{NRPy} implementation and the public \texttt{pySEOBNR} code, we compare \texttt{SEBOB} and other state-of-the-art effective-one-body (EOB) models to all 607 quasi-circular, aligned-spin black hole binary, numerical-relativity (NR) waveforms from the \texttt{SXS} catalog (\texttt{v2025.0.10} release~\cite{Scheel:2025jct}). We then test the key assumptions underlying the BOB merger--ringdown approximation and, finally, benchmark the computational efficiency of our implementation.

Throughout, waveform accuracy is quantified by the noise-free mismatch of the dominant $(2,2)$ strain mode,
\begin{equation}
    \mathcal{M}(h_1,h_2)=1-\frac{|\langle h_1|h_2\rangle|}
    {\sqrt{\langle h_1|h_1\rangle\langle h_2|h_2\rangle}},
    \label{eqn:mismatch_definition}
\end{equation}
with  inner product
\begin{equation}
    \langle h_1|h_2\rangle=\int\!\tilde h_1^{*}(f)\,\tilde h_2(f)\,\mathrm{d}f,
    \label{eqn:inner_product}
\end{equation}
where $\tilde h(f)$ is the Fourier transform of the real part of the time-domain strain $h(t)$. The mismatches are calculated using the open source python package \texttt{PyART}(https://github.com/RoxGamba/PyART).
\subsection{\label{subsec:numerical}Numerical Consistency}

\begin{figure}[ht]
    \centering
    \includegraphics[width=0.48\textwidth]{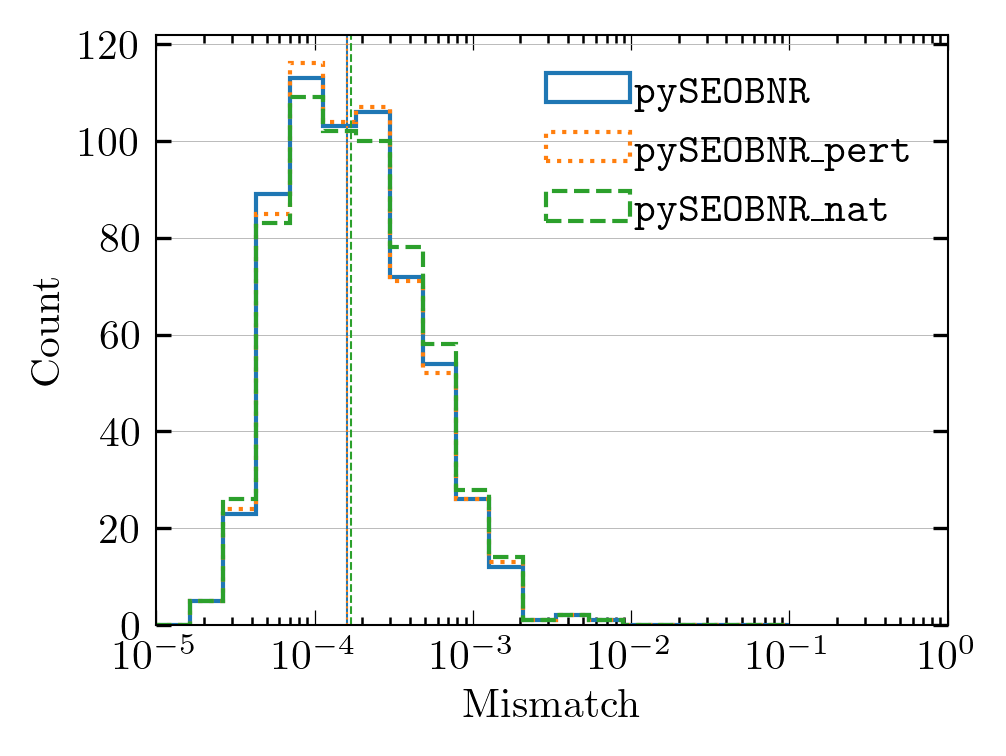}
    \caption{Mismatch histograms between the public \texttt{pySEOBNR} implementation and the \texttt{SXS} catalog. The orange curve (\texttt{pyseobnr\_pert}) shows round-off-level perturbations; the green curve (\texttt{pyseobnr\_nat}) adopts natural-spline boundary conditions. Vertical lines mark median mismatches. The green curve has a slightly higher median mismatch than round-off perturbation indicating a small number of cases where the spline boundary conditions affect the accuracy of the calibration.}
    \label{fig:nat_vs_pip}
\end{figure}

\begin{figure}[ht]
    \centering
    \includegraphics[width=0.48\textwidth]{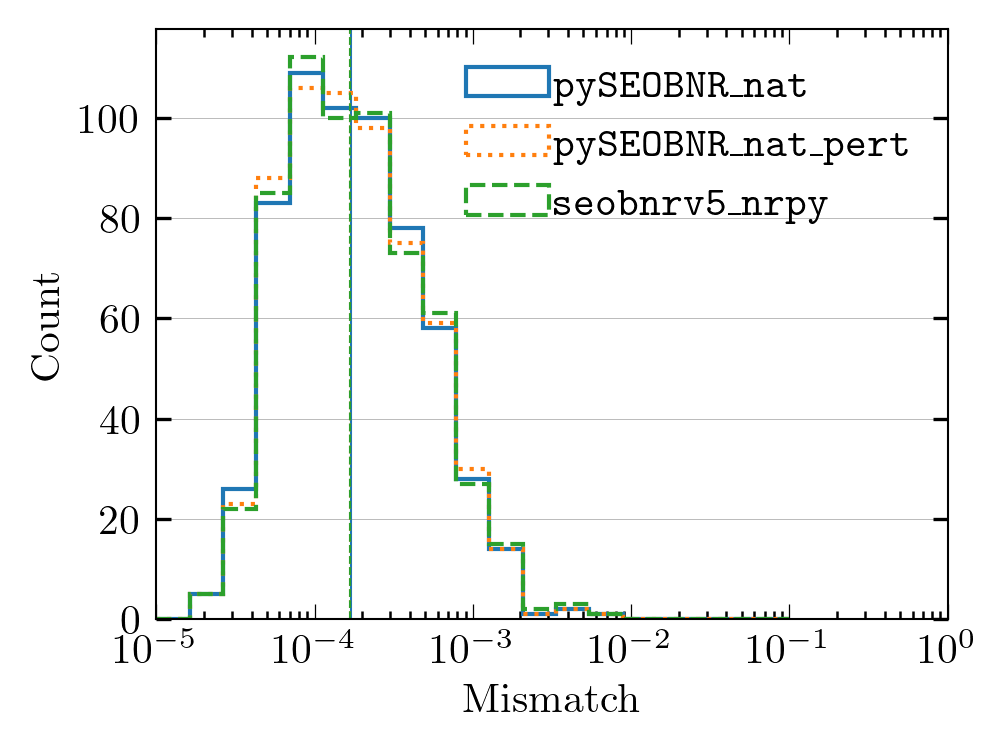}
    \caption{Mismatch histograms comparing the \texttt{NRPy} implementation of \texttt{SEOBNRv5} with \texttt{pyseobnr\_nat} against the SXS catalog. The orange curve (\texttt{pyseobnr\_nat\_pert}) is \texttt{pyseobnr\_nat} with round-off-level perturbations. The \texttt{NRPy} implementation (green) is consistent with the \texttt{pyseobnr\_nat} version due to the mismatches overlapping within the binwidth compared to round-off perturbations.}
    \label{fig:v5nrpy_vs_nat}
\end{figure}

A known implementation difference between \texttt{NRPy} and \texttt{pySEOBNR} is the choice of cubic-spline boundary conditions, which in turn affects non-quasi-circular (NQC) coefficients when the peak strain time is not attained by the trajectory. To isolate purely numerical effects we consider two modified \texttt{pySEOBNR} variants:
\begin{enumerate}
    \item \texttt{pyseobnr\_pert}, which applies random $10^{-15}$-level perturbations to the mass ratio and spins, and
    \item \texttt{pyseobnr\_nat}, which replaces all interpolations with natural splines \footnote{Natural splines set the second derivative to zero at each endpoint, whereas the public code employs ``not-a-knot'' conditions.}.
\end{enumerate}
Figure~\ref{fig:nat_vs_pip} shows that altering the interpolation boundary conditions induces slightly larger mismatches than round-off perturbations.
By contrast, Figure~\ref{fig:v5nrpy_vs_nat} demonstrates round-off-level agreement between \texttt{NRPy} and \texttt{pyseobnr\_nat}, confirming that our code faithfully reproduces the core \texttt{SEOBNRv5} dynamics.
In Table \ref{tab:eob_stops}, we state the typical frequency of cases when the estimated peak time is not encountered. We profile stop conditions for a random combination of 5000 mass ratios, spins, and start frequencies to estimate how often this can be encountered during parameter estimation. We note that NQC conditions evaluated at the endpoint of the trajectory in 10\% of the random sample and 3\% of SXS simulations.
\begin{table}[ht]
    \caption{\label{tab:eob_stops}Frequency of ODE stop conditions for the \texttt{SEOBNRv5} waveform model. The conditions are calculated over 5000 randomly sampled points with $1 \leq q \leq 20$, $-0.999\leq\chi_i\leq.999$, and $5\times 10^{-3}\leq \Omega_{\mathrm{start}} \leq 1.5\times 10^{-2}$. The frequency for the 607 quasi-circular, spin-aligned, black hole binaries simulated by the SXS catalog is given in brackets. For each stop condition, we specify how often the ODE was terminated before remnant ISCO and/or $t_{22}^{peak}$ crossing. The 2 non ISCO crossings when $r < r_{\mathrm{stop}} < r_{\mathrm{ISCO}}$, happens because $dt_{\mathrm{ODE}} \approx .1 \equiv dt_{\mathrm{sample}}$ at last time leading to exclusion of $r_{\mathrm{ISCO}}$ in the fine sampling.}
    \begin{ruledtabular}
    \begin{tabular}{cccc}
    \textbf{Condition}&\textbf{Total}&\textbf{ISCO}&\textbf{Peak}\\
    \hline
    $\Omega$ peak & 3220(369) & 78(0) & 530(20)\\
    $p_{r_{*}}$ peak & 1080(103) & 0(0) & 0(0)\\
    $\dot{r} > 0$ & 6(0) & 0(0) & 0(0)\\
    $r < r_{\mathrm{stop}}$ & 684(132) & 2(0) & 0(0)\\
    $r < 3 $ and $\Omega > 1$ & 10(3) & 0(0) & 0(0)\\
    \end{tabular}
    \end{ruledtabular}
\end{table}
\subsection{\label{subsec:accuracy}Accuracy of \texttt{SEBOB}}

\begin{figure}[ht]
    \centering
    \includegraphics[width=0.48\textwidth]{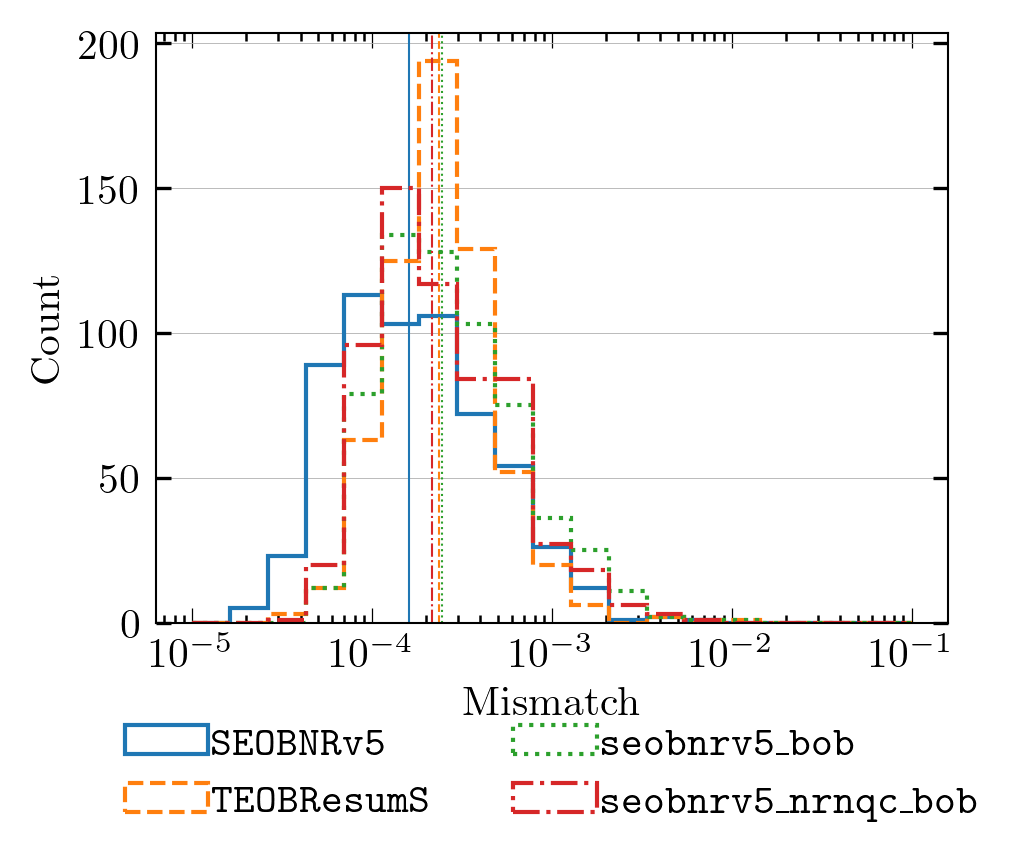}
    \caption{Mismatch histograms comparing state-of-the-art EOB models, including \texttt{SEOBNRv5}, \texttt{TEOBResumS-GIOTTO}~\cite{Nagar:2020pcj}, and the two \texttt{SEBOB} variants that differ only in their NQC prescriptions against the \texttt{SXS} catalog. All models achieve median mismatches $\mathcal{M}\!\approx\!2\times 10^{-4}$.}
    \label{fig:eob_mismatch_analysis}
\end{figure}

\begin{figure}[ht]
    \centering
    \includegraphics[width=0.48\textwidth]{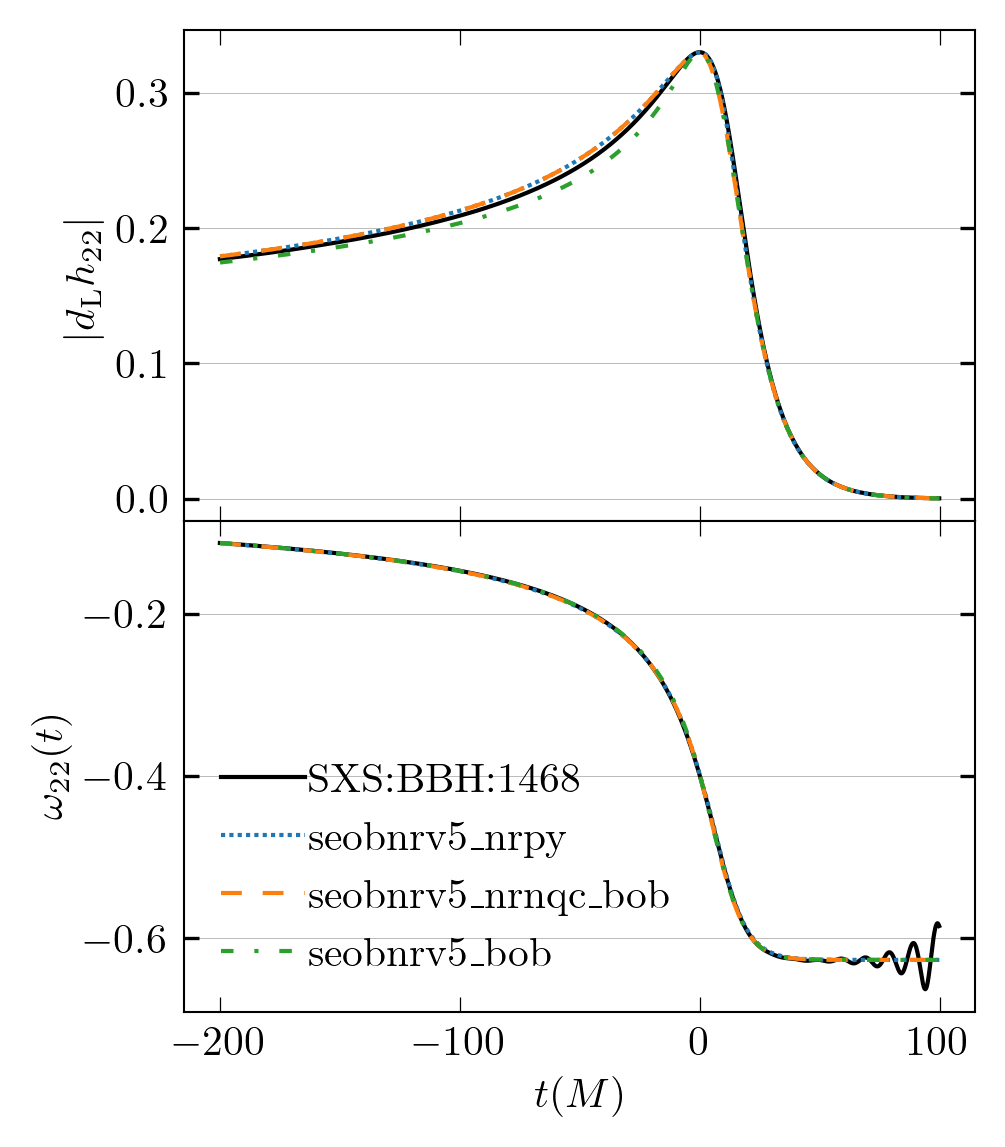}
    \caption{Waveform strain amplitude (top) and frequency (bottom) of the $(2,2)$ mode for the lowest mismatch case (SXS:BBH:1468) ($q = 3, \chi_{1} = -0.599, \chi_{2} = -0.399$), and the corresponding \texttt{seobnrv5\_nrpy}, \texttt{seobnrv5\_nrnqc\_bob}, and \texttt{seobnrv5\_bob} quantities assuming the same system parameters. We note that the BOB-derived NQC corrections provide an accurate description of the frequency before and after the merger.}
    \label{fig:worst_case_v5_nrnqc_bob_v5nrpy_2480}
\end{figure}

\begin{figure}[ht]
    \centering
    \includegraphics[width=0.48\textwidth]{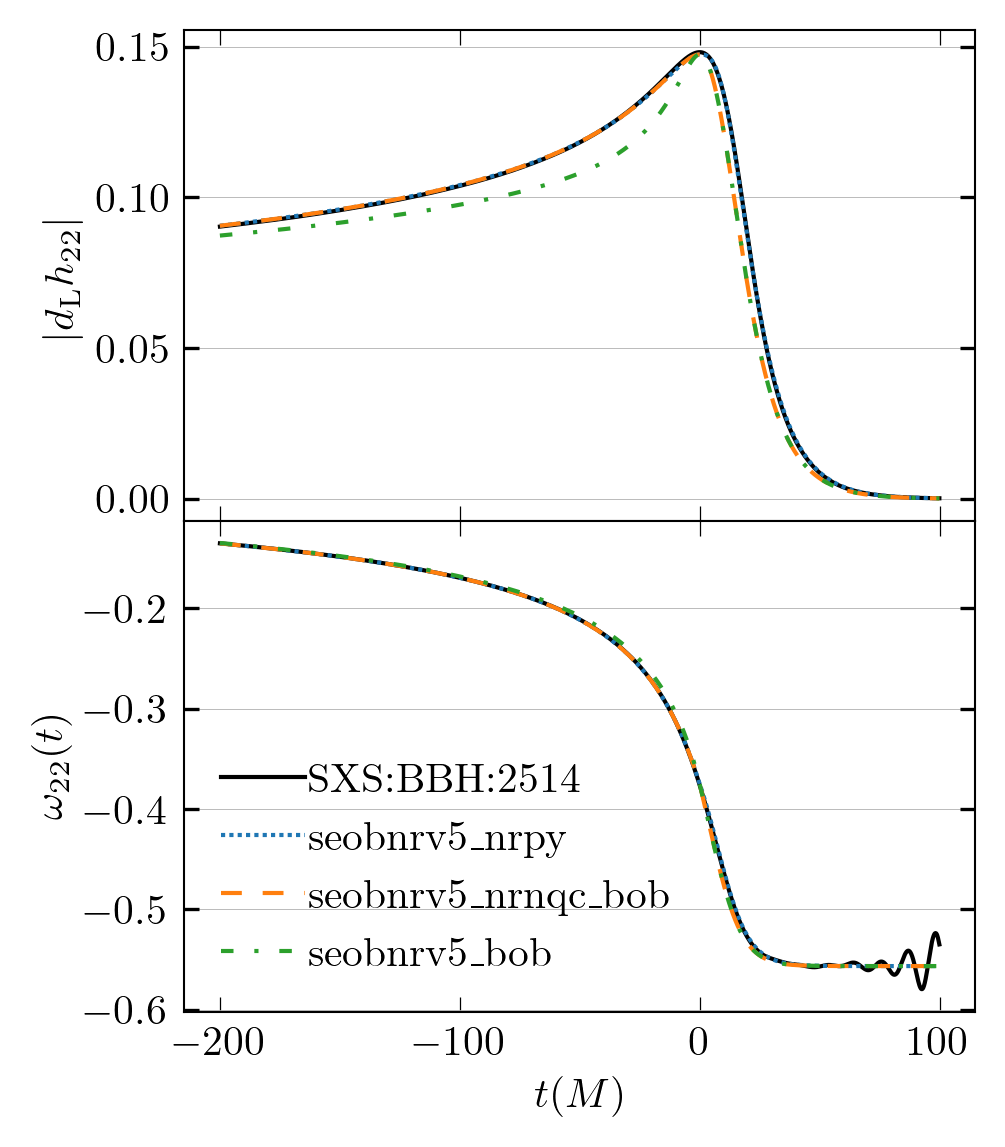}
    \caption{Waveform strain amplitude (top) and frequency (bottom) of the $(2,2)$ mode for a high-mismatch case (SXS:BBH:2514) ($q = 8, \chi_{1} = 0.599, \chi_{2} = 0.599$), and the corresponding \texttt{seobnrv5\_nrpy}, \texttt{seobnrv5\_nrnqc\_bob}, and \texttt{seobnrv5\_bob} quantities assuming the same system parameters. We note the significant loss in amplitude accuracy prior to merger for this system when BOB is used to compute NQC derivatives.}
    \label{fig:worst_case_v5bob_v5nrpy_2514}
\end{figure}

Figure~\ref{fig:eob_mismatch_analysis} compares mismatches against the SXS catalog for all publicly available EOB models. Both \texttt{seobnrv5\_bob} and \texttt{seobnrv5\_nrnqc\_bob} are competitive with existing models, with a small loss in median accuracy due to the replacement of NR-fitted information in the merger-ringdown regime with BOB. 
To highlight how BOB-modelled merger-ringdown and NQC corrections manifest in the waveform, we show the waveform strain amplitude (top) and frequency (bottom) of the $(2,2)$ mode for the the case where BOB is least accurate against SXS compared to \texttt{SEOBNRv5} (Figure~\ref{fig:worst_case_v5_nrnqc_bob_v5nrpy_2480}) and the case where BOB is most accurate against SXS compared to \texttt{SEOBNRv5} (Figure~\ref{fig:worst_case_v5bob_v5nrpy_2514}). Comparing both cases, we find that BOB provides a more accurate description of the frequency evolution compared to the amplitude, even when used as a replacement for NR-fitted NQC corrections. Particularly, in Figure~\ref{fig:worst_case_v5bob_v5nrpy_2514}, we note that \texttt{seobnrv5\_bob} poorly models the curvature (second derivative) of the amplitude close to the peak. Since BOB uses the assumption of adiabaticity ($|h| \approx |\psi_4|\omega^2$, i.e $|\psi_4|$ is slowly varying), the variation of the strain amplitude is not described accurately if the variation in the $\psi_4$ amplitude is non-negligible. In the subsequent discussion, we will assess the validity and impact of the BOB postulates and assumptions.

\subsection{Validity of BOB}

The BOB derivation assumes
(i) a common phase $\phi$ for $\psi_4$, the news $\mathcal{N}$, and the strain $h$,
\begin{equation}
    \phi_{\psi_4}\simeq\phi_{\mathcal{N}}\simeq\phi_h\equiv\phi,
    \label{eqn:common_phase}
\end{equation}
and
(ii) an adiabatically varying amplitude,
\begin{equation}
    |h|\simeq\frac{|\mathcal{N}|}{\omega}\simeq\frac{|\psi_4|}{\omega^2},
    \label{eqn:adiabatic_amplitude}
\end{equation}
where $\omega\equiv \dot\phi$.
To quantify departures from these postulates we compute four root-mean-squared errors (RMSEs) over the interval $[t_0,t_0+20M]$ following peak strain~$t_0$:
\begin{align}
    \epsilon_1 & =\mathrm{RMSE}\!\left(\alpha\,\frac{\mathrm{d}\omega^2}{\mathrm{d}t}, |\mathcal{N}|^2\right);\quad
    \alpha =\left.\left(\frac{|\mathcal{N}|^2}{\mathrm{d}\omega^2/\mathrm{d}t}\right)\right|_{t_0},
    \label{eqn:epsilon_1}                                                                                        \\[2pt]
    \epsilon_2 & =\mathrm{RMSE}\!\bigl(A(t),A_p\,\mathrm{sech}\!\bigl[(t\!-\!t_p)/\tau\bigr]\bigr),
    \label{eqn:epsilon_2}                                                                                        \\[2pt]
    \epsilon_3 & =\mathrm{RMSE}\!\bigl(\omega_{\psi_4},\omega_h\bigr),
    \label{eqn:epsilon_3}                                                                                        \\[2pt]
    \epsilon_4 & =\mathrm{RMSE}\!\bigl(|h|,|\psi_4|/\omega^2\bigr).
    \label{eqn:epsilon_4}
\end{align}

\begin{figure*}[t]
    \centering
    \includegraphics[width=\textwidth]{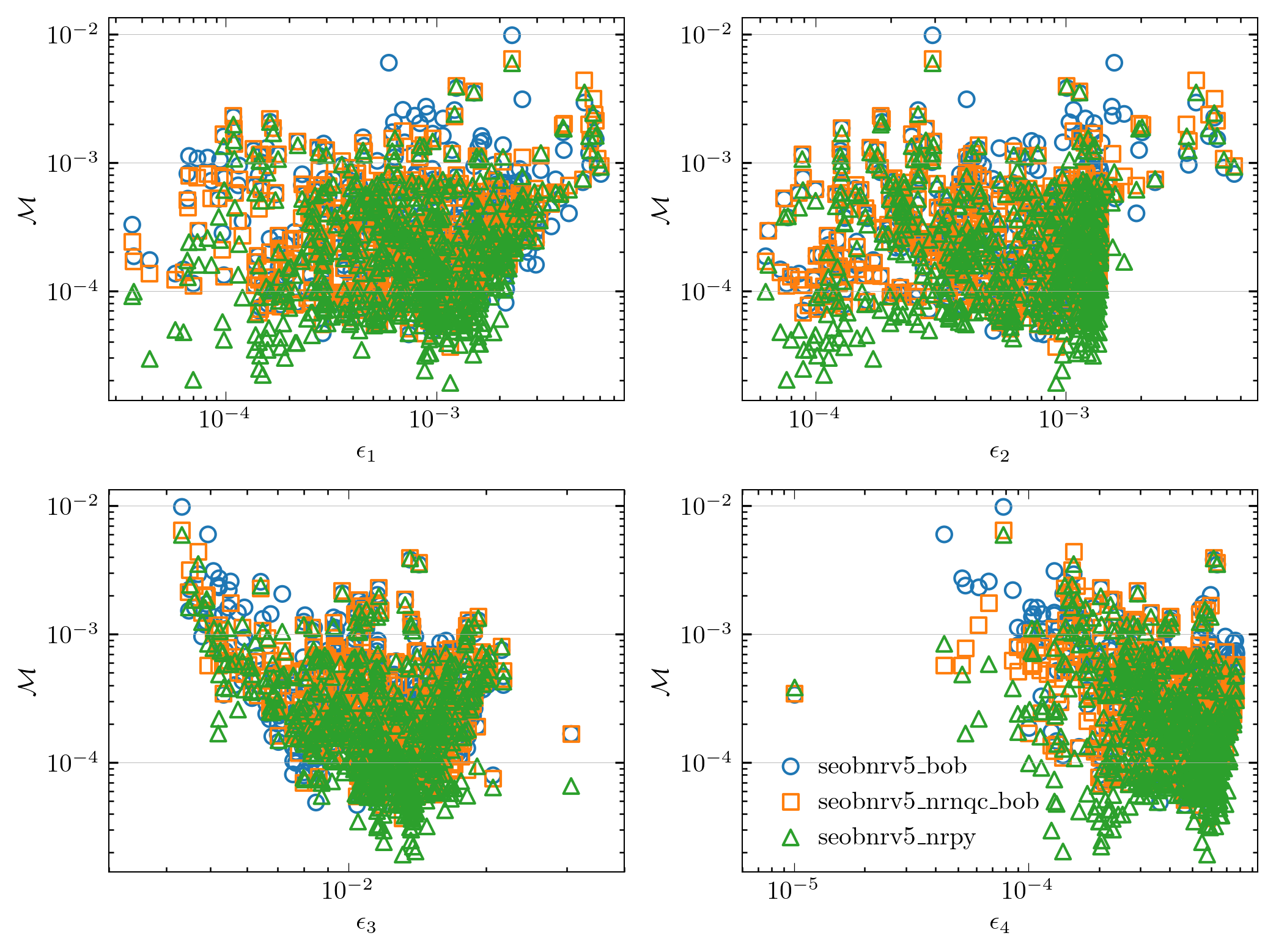}
    \caption{Scatter plots of mismatch versus the four error metrics $\epsilon_{1\text{--}4}$ defined in Eqs.~\eqref{eqn:epsilon_1}--\eqref{eqn:epsilon_4}. The BOB assumptions are approximately satisfied for the catalog, with only weak correlations to the mismatch.}
    \label{fig:correlation_bob_assumptions_v5}
\end{figure*}

Figure~\ref{fig:correlation_bob_assumptions_v5} shows that BOB's core relations hold to good accuracy for quasi-circular, aligned-spin binaries; the largest mismatches coincide with elevated $\epsilon_1$ and $\epsilon_2$. Improvements will therefore focus on the news--frequency linkage and $\psi_4$ amplitude model \cite{Anuj:2025}.

\subsection{NQC Derivatives from BOB}

\begin{figure*}[ht]
    \centering
    \includegraphics[width=\textwidth]{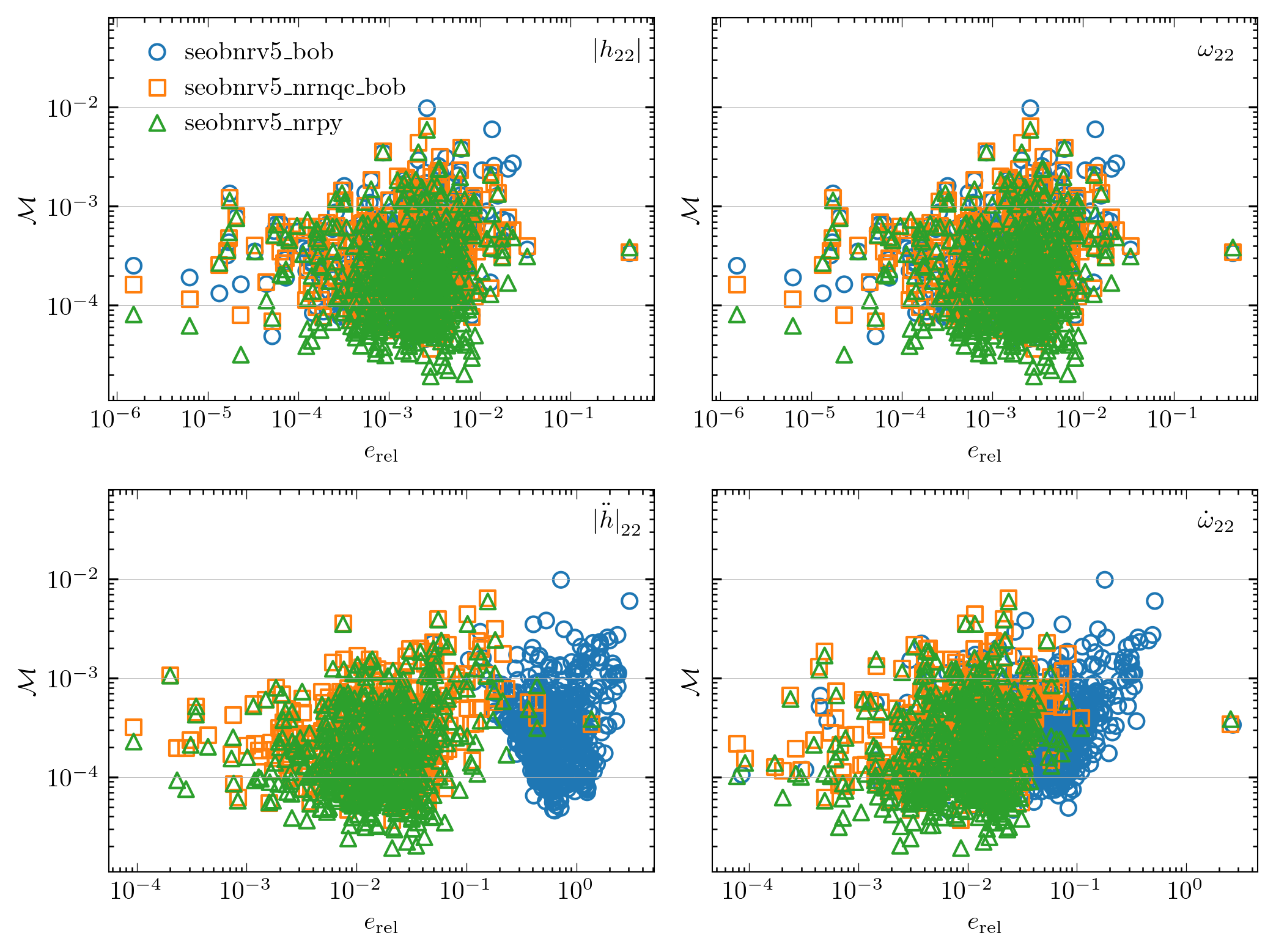}
    \caption{Relative error $e_{\mathrm{rel}}$ of the peak amplitude and frequency fits used in \texttt{SEOBNRv5} (above) and the fits for the amplitude second derivative and frequency derivative provided as fits in \texttt{SEOBNRv5} or derived from BOB NQC parameters (below) versus waveform mismatch against the SXS catalog. Each point corresponds to a waveform in the SXS catalog. The relative errors for the NQC parameters are calculated against their corresponding SXS value at the peak strain time. We again note the overall lack of correlation between these errors and the mismatch. The frequency derivative is predicted to within $\sim4\%$ on average, while the amplitude's second derivative is less accurate, explaining part of the residual mismatch~\cite{Note2}.}
    \label{fig:correlation_bob_nqc_v5_mm}
\end{figure*}

Unlike standard \texttt{SEOBNRv5}, which uses NR-fitted derivatives at peak strain, \texttt{seobnrv5\_bob} computes the second derivative of the amplitude and the first derivative of the frequency analytically from BOB. Their accuracy is assessed in Figure~\ref{fig:correlation_bob_nqc_v5_mm} \footnote{The highest error case in the amplitude and frequency fits corresponds to \texttt{SXS:BBH:1110} ($q=7, \chi_1=3\times10^{-6}, \chi_2=-2\times10^{-7}$), which was deprecated during final development of the paper.\label{footnote:1110deprecation}}. The quality of the NR fits and the BOB-derived quantities are measured by relative error,
\begin{equation}
    e_{\mathrm{rel}}(y_{\mathrm{NR}},y_{\mathrm{pred}})=\frac{|y_{\mathrm{NR}}-y_{\mathrm{pred}}|}{|y_{\mathrm{NR}}|},
\end{equation}
where $y_{\mathrm{NR}}$ is the value of the quantity obtained at the time of the peak strain directly from waveforms in the SXS catalog and $y_{\mathrm{pred}}$ is the predicted quantity either from the fits given in \texttt{SEOBNRv5} or derived from BOB. 

\begin{figure}[ht]
    \centering
    \includegraphics[width=0.48\textwidth]{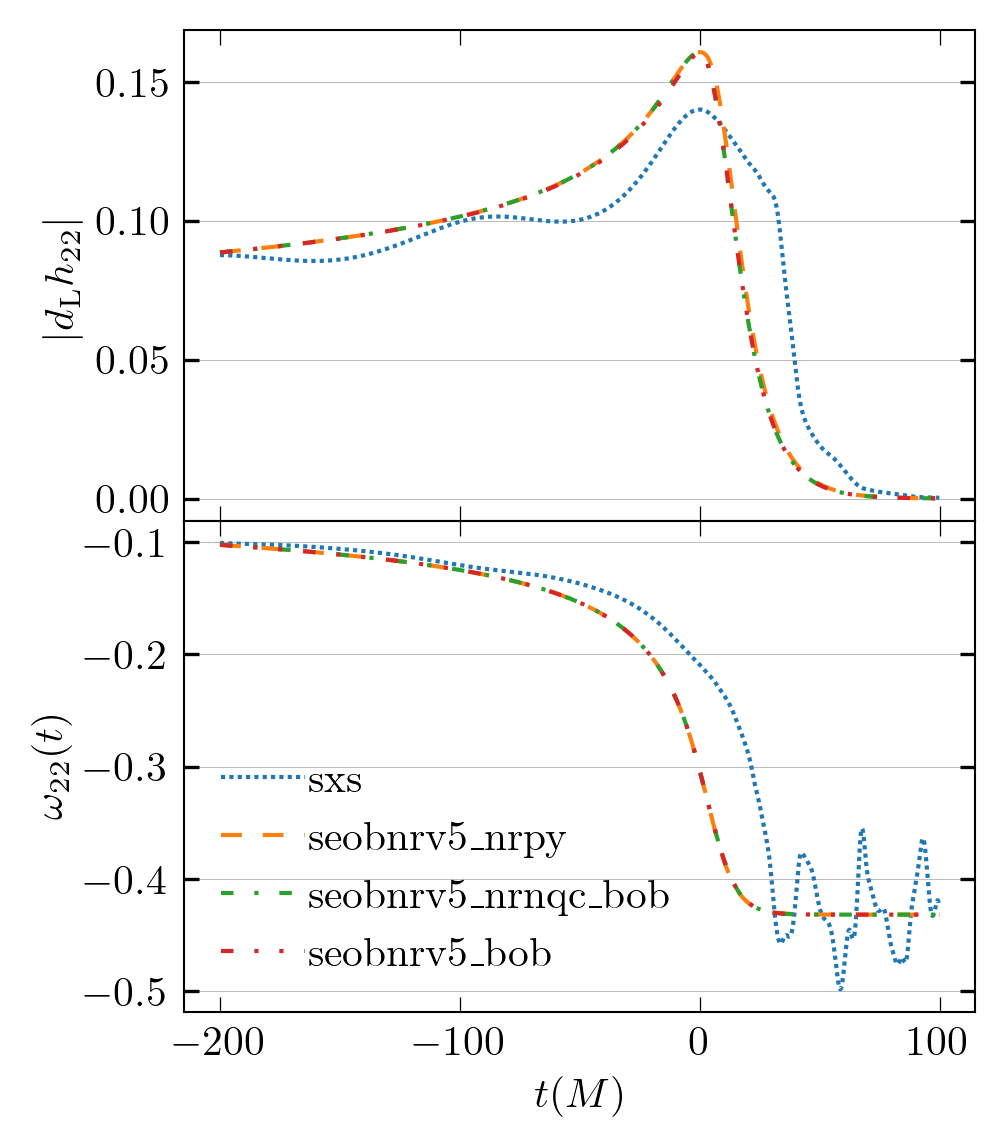}
    \caption{Waveform strain amplitude (top) and frequency (bottom) of the $(2,2)$ mode for the outlying case (SXS:BBH:1110) ($q = 7, \chi_{1} = 3\times10^{-6}, \chi_{2} = -2\times10^{-7}$)~\cite{Note2}, and the corresponding \texttt{seobnrv5\_nrpy}, \texttt{seobnrv5\_bob}, and \texttt{seobnrv5\_nrnqc\_bob} waveforms assuming the same system parameters. Note that all three models are much more consistent with each other despite the different degrees of NR tuning. This comparison can be particularly helpful in detecting anomalous behavior in large waveform catalogs.}
    \label{fig:sxs_v5nrpy_1110}
\end{figure}

The frequency derivative is reproduced to $\sim4\%$, close to the $\sim1\%$ NR-fit uncertainty, whereas the amplitude's second derivative shows $\mathcal{O}(1)$ relative error. This deficiency stems from the assumption of a slowly varying $\psi_4$ amplitude. However, even these sizeable parameter errors translate into only mild increases in mismatch; this is evidence that BOB captures the dominant merger-ringdown physics despite its simplified assumptions. We note that these errors can be decreased significantly by formulating BOB to describe the news, rather than $\psi_4$, and by more accurately calculating the amplitude without assuming adiabaticity, as will be shown in an upcoming paper \cite{Anuj:2025}. Additionally, we note that the single outlying case in the amplitude and frequency fits, as seen in the bottom panels of Figure~\ref{fig:correlation_bob_assumptions_v5} and three of the panels in Figure~\ref{fig:correlation_bob_nqc_v5_mm}, corresponds to \texttt{SXS:BBH:1110} ($q=7, \chi_1=3\times10^{-6}, \chi_2=-2\times10^{-7}$), where the \texttt{SXS} waveform has particularly noisy behavior in the vicinity of merger, as demonstrated in Figure~\ref{fig:sxs_v5nrpy_1110}~\cite{Note2}.

\subsection{\label{subsec:efficiency}Efficiency}

\begin{figure}[ht]
    \centering
    \includegraphics[width=0.48\textwidth]{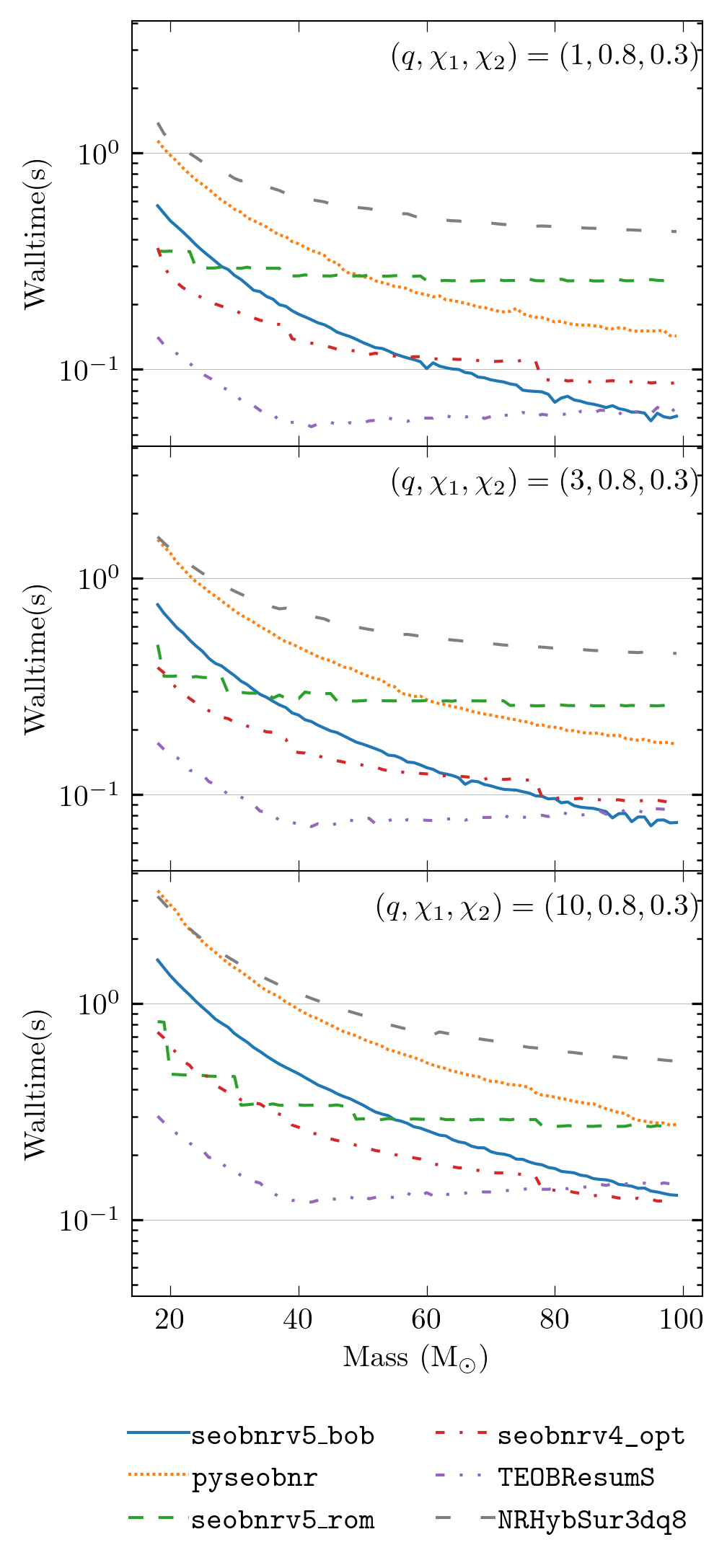}
    \caption{Wall-times for the \texttt{NRPy} implementation of \texttt{seobnrv5\_bob} compared with leading time-domain generators. Times are measured at a 10\,Hz start frequency for $q=\{1,3,10\}$ and aligned spins $(\chi_1,\chi_2)=(0.8,0.3)$}.
    \label{fig:spin_aligned_timing}
\end{figure}

Because \texttt{NRPy} generates highly optimised C code, we anticipate that our implementation leads to faster waveform generation than the native \texttt{pySEOBNR} implementation. Since \texttt{pySEOBNR} only implements a CSE optimized Hamiltonian, the equivalently optimized flux terms in our implementation should significantly reduce computational time in the ODE right-hand-sides. Figure~\ref{fig:spin_aligned_timing} shows that \texttt{seobnrv5\_bob} is $\sim3$ times faster than \texttt{pySEOBNR} for $q=[1,3]$ cases, and is competitive with other state-of-the-art waveform models. The cases shown in Figure~\ref{fig:spin_aligned_timing} are a representative comparison of walltimes, since we found that changing the spin configurations did not affect the relative performance of the waveform models. All timings exclude I/O and post-processing overhead. \texttt{seobnrv4\_opt}~\cite{Devine:2016ovp}, \texttt{NRHybSur3dq8}~\cite{Varma:2018mmi} and \texttt{SEOBNRv5\_ROM} are timed using their \texttt{LALSuite}; \texttt{TEOBResumS-GIOTTO} uses its standalone C code without post-adiabatic speed-ups. One-time \texttt{Cython} compilation costs for \texttt{pySEOBNR} are omitted. In all cases, the walltime is computed by averaging over the walltime of 100 waveform generations. The computations were done using the Thorny Flat cluster at West Virginia University equipped with the Intel® Xeon® Gold 6138 Processor.

%% file: Sections/conclusions.tex
\section{\label{sec:conclusion}Conclusion}
We have introduced \texttt{SEBOB}, a hybrid waveform framework that combines an Effective-One-Body (EOB) inspiral with a Backwards-One-Body (BOB) description of the merger--ringdown for aligned-spin binary black holes. Two realizations were constructed atop \texttt{SEOBNRv5HM}: (i) \texttt{seobnrv5\_nrnqc\_bob}, which retains NR-informed non--quasi-circular (NQC) corrections and attaches a BOB-based merger--ringdown, and (ii) \texttt{seobnrv5\_bob}, which uses BOB to supply the NQC targets themselves, reducing reliance on phenomenological fits and enabling $\mathcal{C}^2$ continuity at the attachment time. Both are implemented in the open-source \texttt{NRPy} ecosystem, combining a user-friendly, transparent Python environment and a modularized, efficient C code.

Across a large set of SXS simulations for quasi-circular, aligned-spin binaries, both realizations attain median mismatches (with a flat psd) of $\mathcal{M}\simeq 2\times10^{-4}$ for the dominant $(2,2)$ mode (Fig.~\ref{fig:eob_mismatch_analysis}), demonstrating that an analytically motivated merger--ringdown can maintain state-of-the-art fidelity. Diagnostics identify a localized imperfection in the curvature of the strain amplitude near its peak when NQC corrections are informed by the current $\psi_4$-based version of BOB; by contrast, the frequency evolution is captured to within a few percent, with typical peak $|\dot{\omega}|$ errors of $\sim4\%$ (Fig.~\ref{fig:correlation_bob_nqc_v5_mm}). Catalog-wide tests indicate that the core BOB postulates are satisfied to useful accuracy, with only weak correlations between their measured deviations and the overall mismatch (Fig.~\ref{fig:correlation_bob_assumptions_v5}). Even with the localized amplitude shortfall, total mismatches remain at the few-$10^{-4}$ level.

A key practical outcome is computational performance. With symbolic optimization and comprehensive common--subexpression elimination applied to both the Hamiltonian and radiation--reaction fluxes, the \texttt{NRPy}-generated C implementation achieves $\sim3\times$ shorter wall times than \texttt{pySEOBNR} for equal and moderate mass ratios while remaining competitive with other time-domain generators (Fig.~\ref{fig:spin_aligned_timing}). Since we do not benchmark the time taken to output and process waveforms, a caveat to Fig.~\ref{fig:spin_aligned_timing} is that we do not take into account the time taken to produce a template bank or to perform parameter inference as part of our walltime analysis.

A key limitation of the BOB-informed NQC corrections is a lack of overlap between the EOB and BOB physics in the window where the corrections are necessary. EOB uses PN factorized waveform and fluxes, assumes circularity, and requires late-inspiral corrections to account for highly non-circular orbital motion and the general breakdown of PN close to merger. Upcoming work will explore constructing BOB directly from the news function without the assumption of adiabaticity, which should significantly reduce errors in the amplitude derivatives close to the peak \cite{Anuj:2025}. Within the current paradigm of NQC corrections, this version of BOB can mitigate the need for higher derivatives of NR data to accurately model the amplitude corrections in the late inspiral.

A long-term goal for the \texttt{SEBOB} formalism is to eliminate the remaining direct reliance on NR fittings entering the EOB inspiral and dynamics. EOB models rely on extensive calibration of inspiral waveforms to numerical relativity (NR) waveforms to enhance accuracy. In \texttt{SEOBNRv5}, the underlying waveforms already incorporate NR-derived through NQC corrections and the phenomenological merger-ringdown (see Table~\ref{tab:nr_dof} for a breakdown of pre-calibration NR parameters). A key advantage of \texttt{SEBOB} is that it limits pre-calibration NR information to 4 parameters ($|h|_0,\omega_0,M_f,a_f$), significantly reducing the need for NR to model the merger-ringdown within calibration accuracy. It also reduces the need for extensive calibrations by providing a substantially accurate, physically motivated map from inspiral to merger-ringdown physics. Furthermore, self-consistently modeling non-circularities into the EOB fluxes and inferring remnant properties directly from the EOB inspiral (as suggested in \cite{Buonanno:2007sv}) can potentially reduce the number of NR parameters needed in the inspiral sector. Such improvements would significantly limit the need for NR calibration to a narrow window (between the EOB plunge and the BOB merger) and fewer parameters.

The present study is limited to spin-aligned, quasi-circular binaries and to the $(2,2)$ mode, and it continues to use NR fits for remnant properties. Several targeted extensions are natural: (i) improving the amplitude model by formulating BOB in terms of the news rather than $\psi_4$ and relaxing adiabaticity to sharpen NQC derivatives~\cite{Anuj:2025}; (ii) incorporating higher harmonics and, ultimately, generic precessing spins; (iii) tightening the inspiral-to-remnant mapping to further curtail NR-derived inputs; and (iv) developing an open-source pipeline to perform targeted calibrations to NR. By substantially shrinking the role of phenomenological calibration in the most nonlinear phase while retaining competitive accuracy and speed, \texttt{SEBOB} offers a robust, interpretable, and extensible path toward waveform modeling in the high-SNR era.